\documentclass[useAMS,usenatbib]{mn2e}

\usepackage{graphicx}

\newcommand{\ion}[2]{#1\,{\sc #2}}

\newcommand{\logg}{\log\,g}
\newcommand{\teff}{T_{\rm eff}}

\title[Variability of the PGa star HD\,19400]{
Abundance analysis, spectral variability, and search
for the presence of a magnetic field in the typical PGa star HD\,19400
}

\author[Hubrig et al.]{
S.~Hubrig$^{1}$\thanks{E-mail: shubrig@aip.de},
F.~Castelli$^{2}$,
J.~F.~Gonz\'alez$^{3}$,
T.~A.~Carroll$^{1}$,
I.~Ilyin$^{1}$,
M.~Sch\"oller$^{4}$,
\and
N.~A.~Drake$^{5,6}$,
H.~Korhonen$^{7}$,
M.~Briquet$^{8}$ \\
$^1$ Leibniz-Institut f\"ur Astrophysik Potsdam (AIP), An der Sternwarte 16, 14482 Potsdam, Germany \\
$^2$ Istituto Nazionale di Astrofisica, Osservatorio Astronomico di Trieste, via Tiepolo 11, 34143 Trieste, Italy  \\
$^3$ Instituto de Ciencias Astronomicas, de la Tierra, y del Espacio (ICATE), 5400 San Juan, Argentina \\
$^4$ European Southern Observatory, Karl-Schwarzschild-Str.~2, 85748 Garching bei M\"unchen, Germany \\
$^5$ Sobolev Astronomical Institute, St.~Petersburg State University, Universitetski pr.~28, 198504, St.~Petersburg, Russia \\
$^6$ Observat\'orio Nacional/MCTI, Rua Jos\'e Cristino 77, CEP 20921-400, S\~ao Crist\'ov\~ao, Rio de Janeiro, RJ, Brazil \\
$^7$ Finnish Centre for Astronomy with ESO (FINCA), University of Turku, V\"ais\"al\"antie 20, 21500, Piikki\"o, Finland \\
$^8$ Institut d'Astrophysique et de G\'eophysique, Universit\'e de Li\`ege, All\'ee du 6 Ao\^ut 17, Sart-Tilman, B\^at.\ B5C, 4000, Li\`ege, Belgium
}

\begin{document}

\date{Accepted Received; in original form}

\pagerange{\pageref{firstpage}--\pageref{lastpage}} \pubyear{2014}

\maketitle

\label{firstpage}

\begin{abstract}

The aim of this study is to carry out an abundance determination, to search for spectral variability 
and for the presence of a weak magnetic field in the typical PGa star HD\,19400.
High-resolution, high signal-to-noise HARPS spectropolarimetric observations of HD\,19400 were obtained at three 
different epochs in 2011 and 2013. 
For the first time, we present abundances of various elements determined using an ATLAS12 model, 
including the abundances of a number of elements not analysed by previous studies, such as \ion{Ne}{i}, \ion{Ga}{ii}, and 
\ion{Xe}{ii}. Several lines of \ion{As}{ii} are also present in the spectra of HD\,19400.
To study the variability, we compared 
the behaviour of the line profiles of various elements.
We report on the first detection of anomalous shapes of line profiles belonging
to Mn, and Hg, and the variability of the line profiles belonging to the elements 
Hg, P, Mn, Fe, and Ga.
We suggest that the variability of the line profiles of these elements is caused by their non-uniform surface distribution,
similar to the presence of chemical spots detected in HgMn stars.
The search for the presence of a magnetic field was carried out using the moment technique and the SVD method.
Our measurements of the magnetic field with the moment technique using 22 \ion{Mn}{ii} lines  
indicate the potential existence of a weak variable longitudinal magnetic field on the first epoch.
The SVD method applied to the \ion{Mn}{ii} lines indicates $\left<B_z\right>=-76\pm25$\,G on the first epoch, and
at the same epoch the SVD analysis of the observations
using the \ion{Fe}{ii} lines shows $\left<B_z\right>=-91\pm35$\,G.
The calculated false alarm probability values, 0.008 and 0.003,
respectively, are above the value $10^{-3}$, indicating no detection.
\end{abstract}


\section{Introduction}
\label{sect:intro}

A number of chemically peculiar stars with spectral types B7--B9 exhibit in their 
atmospheres large excesses of P, Mn, Ga, Br, Sr, Y, Zr, 
Rh, Pd, Xe, Pr, Yb, W, Re, Os, Pt, Au, and Hg, and 
underabundances of He, Al, Zn, Ni, and Co (e.g., \citealt{cast2004}). 
These stars are usually called the HgMn stars. 
The aspect of inhomogeneous distribution of some chemical elements over the surface of HgMn stars 
was first discussed by \citet{HubrigMathys1995}. From a survey of HgMn stars in close 
spectroscopic binaries, they
suggested that some chemical elements might be inhomogeneously distributed on the surface, with, 
in particular, preferential concentration of Hg along the equator. 
Recent studies revealed that not only Hg, but also many other elements, most typically Ti, Cr, Fe, 
Mn, Sr, Y, and Pt, are concentrated in spots of diverse size, and different elements exhibit 
different abundance distributions across the stellar surface (e.g.\ \citealt{Hubrig2006a}; 
\citealt{Briquet2010}; \citealt{Mak2011a}; \citealt{Korhonen2013}).
In SB2 systems, the hemispheres of components facing each other usually display low-abundance element spots, or 
no spots at all (e.g.\ \citealt{Hubrig2010}).
Moreover, evolution of the abundance spots of several elements at different time scales was 
discovered in 
a few HgMn stars: 
\citet{Briquet2010} and \citet{Korhonen2013} reported the 
presence of dynamical spot evolution over a couple of weeks for the SB1 system HD\,11753, 
while \citet{Hubrig2010} detected a secular element evolution in the double-lined eclipsing 
binary AR\,Aur.

However, not much is known about the behaviour of different elements in the hotter extension of the HgMn 
stars, the PGa stars, with rich \ion{P}{ii},  \ion{Mn}{ii},  \ion{Ga}{ii}, and  \ion{Hg}{ii} spectra, and effective temperatures 
of about 13\,500\,K and higher (e.g., \citealt{alonso2003}; \citealt{rach2006}).

During our observing run in 2013 July, we  obtained a high-resolution, high signal-to-noise (S/N) polarimetric 
HARPS spectrum of the typical PGa star HD\,19400.
We downloaded two additional polarimetric spectra of this star, obtained 
on two consecutive nights in 2011 December, from the ESO archive.
These spectra were used to carry out an abundance analysis and to investigate whether HD\,19400, 
similar to HgMn stars, exhibits a weak magnetic field 
and an inhomogeneous distribution of various elements over the stellar surface.
Notably, \citet{Maitzen1984} suggested the presence of a magnetic field in 
this star using observations of the $\lambda$5200 feature.
A careful inspection of the spectra acquired on three different 
epochs revealed the presence of anomalous flat-bottom line profiles belonging to the overabundant 
elements Hg and Mn \citep{drake2013}, reminiscent of profile shapes observed in numerous HgMn stars (e.g.,
\citealt{Hubrig2006a,Hubrig2011}; \citealt{Mak2011a}). 
Moreover, these observations revealed the variability of 
line profiles belonging to \ion{Hg}{ii}, \ion{Mn}{ii}, \ion{P}{ii}, \ion{Fe}{ii}, and \ion{Ga}{ii}.
\citet{DommangetNys2002} mention in the 
CCDM catalogue a nearby component at a separation of 0\farcs1 and a position angle of 179$^{\circ}$. However, 
no lines belonging to the secondary were detected in the previous  spectral studies, indicating 
that HD\,19400 can be treated as a single star.
In the following sections, we discuss the results of the abundance determination, the spectral variability detected  
in the lines of certain elements and our search for the presence of a weak magnetic field.

\section{Observations}
\label{sect:obs}

All three spectropolarimetric observations have been obtained with the HARPS polarimeter
(HARPSpol; \citealt{snik2008})
attached to ESO's 3.6\,m telescope (La Silla, Chile).
Two spectropolarimetric observations have been obtained on two consecutive nights on 2011 December 15 and 16, 
and one on 2013 July 19.
The obtained polarimetric observations with a $S/N$ between 500 and 600 in the Stokes~$I$ spectra and a resolving 
power of about $R = 115\,000$
cover the spectral range  3780--6910\,\AA{}, with a small gap between 5259 and 5337\,\AA{}.
Each polarimetric observation consists of several subexposures, 
obtained with different orientations of the quarter-wave retarder plate relative to the 
beam splitter of the circular polarimeter. 
The reduction and calibration of archive spectra was performed
using the HARPS data reduction software available at the ESO headquarter in Germany, 
while the spectra obtained in 2013 July have been reduced using the
pipeline available at the 3.6\,m telescope in Chile. 

To normalise the HARPS spectra to the continuum level, we used the image of the extracted echelle orders.
First, we fit a continuum spline in columns of the image in cross-dispersion direction. 
Each column is fitted in a number of subsequent iterations until it converges to the same upper envelope of 
the continuum level. After each iteration, we analyze the residuals of the fit and make a robust estimation 
of the noise level based upon a statistical test of the symmetric part of the distribution. All pixels whose 
residuals are below the specified sigma clipping level are masked out from the subsequent fit. This way the smooth 
spline function is rejecting all spectral lines below, but leaving the continuum pixels to fit. Once all columns 
are processed, we fit the resulting smoothed curves in the dispersion direction by using the same approach with 
the robust noise estimation from the residuals, but this time rejecting possible outliers above and below the
specified sigma clipping level. As a result, we create a bound surface with continuous first derivatives 
in the columns and rows. We employ a smoothing spline with adaptive optimal regularisation parameters, which 
selects the minimum of the curvature integral of the smoothing spline. As a test for the validity of the 
continuum fit, we check whether the normalised overlapping echelle orders are in good agreement with each other. 
The same is applied to the very broad hydrogen lines, whose wings may span over two or  even three spectral 
orders. The typical mismatch between the red and blue ends of the neighboring orders is well within the statistical 
noise of these orders. The usual procedure to normalise a series of polarimetric observations of the same target,
but with different angles of the retarder, is to create a sum of the individual observations, normalise it to the 
continuum in the way described above, and to use the master normalised image as a template for the individual 
observations: by taking the ratio and fitting a regular spline to it, which then finally defines the 
continuum surface for the individual observations.

\begin{table}
\caption{
Logbook of the HARPS polarimetric observations, including 
the modified Julian date of mid-exposure followed by the 
achieved signal-to-noise ratio.
}
\label{tab:rv}
\centering
\begin{tabular}{lc}
\hline
\hline
\multicolumn{1}{c}{MJD} &
\multicolumn{1}{c}{S/N$_{4500}$} \\
\hline
55910.054 &  820 \\
55911.042 &  760 \\
56492.327 &  470 \\
\hline
\end{tabular}
\end{table}

The Stokes~$I$ and $V$ parameters were derived following the ratio method described by 
\citet{Donati1997}, and null polarisation spectra 
were calculated by combining the subexposures 
in such a way that polarisation cancels out.
These steps ensure that no spurious signals are present in the obtained data (e.g.\ \citealt{Ilyin2012}).
The observing logbook is presented in Table~\ref{tab:rv}, where the first column 
gives the date of observation, followed by 
the S/N ratio per resolution element of the spectra
in the wavelength region around 4500\,\AA{}.

\section{Model parameters and abundances of HD\,19400}
\label{sect:params}

The  starting model parameters of HD\,19400 were
derived  from Str\"omgren photometry. The observed colors 
$(b-y)=-0.066$, $m_1=0.111$, $c_1=0.512$, $\beta=2.708$ were taken
from the \citet{hauck1998} Catalogue\footnote{http://obswww.unige.ch/gcpd/gcpd.html}.
The synthetic colors were taken from the grid computed 
for $[{\rm M}/{\rm H}]=0$ and microturbulent velocity $\xi=0$\,km\,sec$^{-1}$
(\citealt{cast2003}; \citealt{cast2006})\footnote{http://wwwuser.oat.ts.astro.it/castelli/colors/\linebreak[3]uvbybeta.html}.
Zero reddening was adopted for this star, in agreement with
the results from the UVBYLIST code of \citet{moon85}. Observed $c_1$ and $\beta$ 
indices were reproduced by synthetic  indices for model parameters 
$\teff=13\,868\pm150$\,K and $\logg=3.81\pm0.06$, where the errors are
associated with estimated errors of $\pm$0.015\,mag and $\pm$0.005\,mag for the
observed $c_1$ and $\beta$ indices, respectively.    

The parameters from the photometry were adopted for computing an ATLAS9 model
with solar abundances for all the elements and zero microturbulent velocity.
Using the WIDTH code \citep{Kurucz2005}, we derived \ion{Fe}{ii} and \ion{Fe}{iii}
abundances from the equivalent widths of 34 \ion{Fe}{ii} and four \ion{Fe}{iii}
lines. 
The equivalent widths were measured with the SPLOT task of the IRAF package
using the ``e'' option, which integrates the intensity over the line profile.
No \ion{Fe}{i} equivalent widths were measured because
the observed lines are weak and blended. 
The \ion{Fe}{ii} and \ion{Fe}{iii} abundances both satisfied the ionisation 
equilibrium condition and provided a good agreement between 
most of the observed and computed blended \ion{Fe}{i} weak profiles.
  We did not find any
trend of \ion{Fe}{ii} abundances with the excitation potential, indicating
that the adopted temperature is correct. We also did not find any trend
of \ion{Fe}{ii} abundances with equivalent widths, indicating that also the
assumption of zero microturbulent velocity is correct. 
For solar abundances, the model was also able to reproduce the 
Balmer lines, indicating that the adopted gravity is correct. 

The ATLAS9 model was used to derive the abundance for all
those  elements that show  lines in the spectrum.
Whenever possible,  equivalent widths were measured. 
For weak and blended lines and for lines that are blends of 
transitions belonging to the same multiplet, such as
\ion{Mg}{ii} 4481\,\AA{}, \ion{He}{i} lines, and most \ion{O}{i}
lines, we derived the abundance from the line profiles.
The synthetic spectrum was also used to determine upper
abundance limits from those lines predicted for solar
abundances, but not observed.

The SYNTHE code \citep{Kurucz2005}, together with line lists  based mostly on 
Kurucz's data \citep{cast2004,cast2010,Kurucz2011,yuce2011} and 
including also data taken from the NIST database 
(version 5)\footnote{http://www.nist.gov/pml/data/asd.cfm} were used to 
compute the synthetic spectrum. 
The synthetic spectrum was broadened both for a Gauss profile corresponding to
the 115\,000 resolving power of the HARPS instrument and for a 
rotational velocity $v\,\sin\,i=32$\,km\,s$^{-1}$. This value was derived from 
the comparison of the observed and computed
profile of several lines.
We estimate an uncertainty of the order of 0.5\,km\,s$^{-1}$ for
this choice.

\begin{figure}
\centering
\includegraphics[angle=90,width=0.45\textwidth]{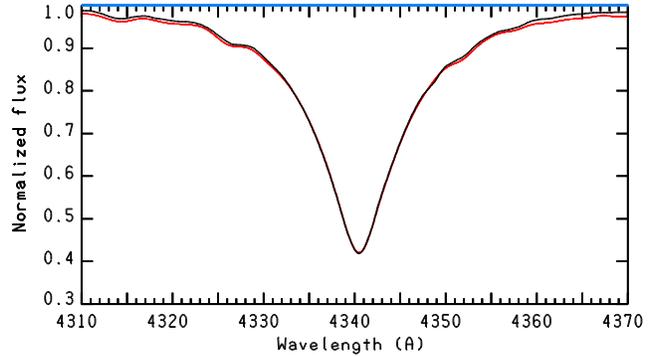}
\caption{Comparison of the H$_{\gamma}$ profile observed in the low-resolution
FORS\,1 spectrum (black) with that computed using the ATLAS12 model with parameters
$\teff=13\,500$\,K and $\logg=3.9$ (red in the online version).}
\label{fig:hydr}
\end{figure}

Once all abundances were determined in this way, we computed an ATLAS12
model \citep{Kurucz2005} for the individual abundances having the same parameters as for
the ATLAS9 model. However, the new ATLAS12 model  did not reproduce 
neither the \ion{Fe}{ii}-\ion{Fe}{iii} equilibrium  nor the hydrogen lines.
In fact, the non-solar abundances of several elements, and in particular
 the helium underabundance, altered the model structure in a consistent way.
We therefore searched for the ATLAS12 model 
adequate to assure the \ion{Fe}{ii}-\ion{Fe}{iii} ionisation equilibrium and 
the Balmer lines reproducibility. We found that the ATLAS12 model 
with parameters $\teff=13\,500$\,K and $\logg=3.9$ met these requirements.
The comparison between the computed H$_{\gamma}$ profile and the H$_{\gamma}$ profile observed in the 
FORS\,1 spectrum at a resolution of $\sim$2000 on 2003 August 1 (ESO Prg.\ 71.D-0308(A)) 
is presented in Fig.~\ref{fig:hydr}.

\begin{table*}
\caption{Abundances $\log$(N$_{\rm elem}$/N$_{\rm tot}$) for HD\,19400.
For each element listed in the first column, we present the abundance computed
using the ATLAS12 model in the second column.
In parentheses is the number of lines adopted to derive the abundance for a given ion.
Blended lines were counted only once. 
In the third column, we list the deviations from solar abundances \citep{asplund2009} presented 
in column 4. The last column gives the abundances derived by  \citet{alonso2003}.
}
\label{tab:abundance}
\centering
\begin{tabular}{lccrccccccccccccccc}
\hline
\hline\noalign{\smallskip}
\multicolumn{1}{c}{Element}&
\multicolumn{1}{c}{HD\,19400}&
\multicolumn{1}{c}{Star-Sun}&
\multicolumn{1}{c}{Sun}&
\multicolumn{1}{c}{\citet{alonso2003}}\\
& [13500\,K,3.9,ATLAS12]& & & [13350\,K,3.76,ATLAS9]\\
\hline\noalign{\smallskip}
\ion{He}{i} & $-$2.17$\pm$0.08: (14) & [$-$1.11]: & $-$1.05 & $-$1.52 \\
\ion{C}{ii} & $-$4.12$\pm$0.02 (4) &[$-$0.51]  & $-$3.61&$-$3.52$\pm$0.28\\ 
\ion{O}{i}  & $-$3.90 (2)  &  [$-$0.55]  & $-$3.35 & $-$3.31$\pm$0.07\\
\ion{Ne}{i} & $-$3.77$\pm$0.07 (6) & [$+$0.34]  & $-$4.11 & \\
\ion{Na}{i} & $-$5.71 (2)  &  [$+$0.09] & $-$5.80\\
\ion{Mg}{ii} & $-$5.06 (4) & [$-$0.62] & $-$4.44 &$-$4.65$\pm$0.30\\
\ion{Al}{ii} &$\le$$-$6.77 (2) &$\le$[$-$1.18]  & $-$5.59 & \\
\ion{Si}{ii} & $-$4.36$\pm$0.17 (10) & [$-$0.17]    & $-$4.53&$-$4.28$\pm$0.31\\
\ion{Si}{iii} & $-$4.37$\pm$0.02 (2) & [$-$0.16]    & $-$4.53&   \\
\ion{P}{ii} & $-$4.26$\pm$0.15 (33) & [$+$2.28]  & $-$6.63 &$-$5.95$\pm$0.16\\
\ion{P}{iii}& $-$4.44$\pm$0.09 (4) & [$+$2.19]  &$-$6.63\\
\ion{S}{ii} &$-$5.82   (1)        & [$-$0.90]  & $-$4.90&$-$5.07$\pm$0.47\\
\ion{Ca}{ii} & $-$5.50: (1) & [$+$0.20]: & $-$5.70& \\
\ion{Ti}{ii} & $-$6.35$\pm$0.07 (9) &[$+$0.74]  & $-$7.09 &$-$5.69$\pm$0.37\\
\ion{Cr}{ii} & $-$6.24$\pm$0.09 (5) &[$+$0.16] & $-$6.40 &$-$5.45$\pm$0.42\\
\ion{Mn}{ii} & $-$4.94$\pm$0.18 (6) &[$+$1.67] & $-$6.61 & $-$4.57$\pm$0.33\\
\ion{Fe}{ii} & $-$3.79$\pm$0.14 (35) & [$+$0.75]     & $-$4.54&$-$3.75$\pm$0.31\\
\ion{Fe}{iii} & $-$3.82$\pm$0.10 (4) &[$+$0.72]     & $-$4.54&$-$3.67$\pm$0.30\\
\ion{Ni}{ii} & $-$5.84 (1)  &  [$-$0.02]  & $-$5.82 & $-$5.51$\pm$0.34\\
\ion{Ga}{ii} & $-$5.19$\pm$0.17 (12) &  [$+$3.81]  & $-$9.00 &\\
\ion{Sr}{ii} & $-$9.07 (1) &  [$+$0.10]  & $-$9.17 & $-$6.95$\pm$0.38\\
\ion{Xe}{ii} &$-$4.65$\pm$0.17 (6) & [$+$5.15]  & $-$9.80\\
\ion{Hg}{ii} & $-$6.16$\pm$0.13 (3) & [$+$4.71]    &$-$10.87&$-$4.43\\
\hline\noalign{\smallskip}
\end{tabular}
\end{table*}

The abundances $\log$(N$_{\rm elem}$/N$_{\rm tot}$) of HD\,19400 derived from 
the ATLAS12 model 
either from equivalent widths or line profiles are listed in Table~\ref{tab:abundance},
together with the solar abundances taken from \citet{asplund2009}.
Similar to previous spectroscopical studies of HD\,19400, no lines belonging to the secondary were detected
in the three HARPS spectra.
All the lines and atomic data used for the abundance analysis are
listed in Table~\ref{tab:line_abundance}.

\begin{table*}
\caption[ ]{
Line by line abundances of HD\,19400 from the ATLAS12 model with parameters
$\teff=13\,500$\,K, $\logg=3.9$. In the second and third columns, we give the oscillator strength with the 
corresponding data base source. The low excitation potential is listed in column 4, followed by the equivalent width and 
the derived abundance. For a number of lines, the abundance was derived from line profiles. The full table is available online.
}
\label{tab:line_abundance}
\font\grande=cmr7
\grande
\centering
\begin{tabular}{lllrrcl}
\hline\noalign{\smallskip}
\multicolumn{7}{c}{HD\,19400[13500,3.9,ATLAS12]}
\\
\hline\noalign{\smallskip}
\multicolumn{1}{c}{$\lambda$(\AA{})} &
\multicolumn{1}{c}{$\log\,gf$}&
\multicolumn{1}{c}{Ref.$^{a}$}&
\multicolumn{1}{c}{$\chi_{\rm low}$}&
\multicolumn{1}{c}{W(m{\AA{}})}&
\multicolumn{1}{c}{$\log$(N$_{\rm elem}$/N$_{\rm tot}$)}&
\multicolumn{1}{l}{Remarks}
\\
\hline\noalign{\smallskip}
\multicolumn{7}{c}{$\log($N(\ion{He}{i})/N$_{\rm tot}) = -2.17\pm0.08$:}\\
\hline\noalign{\smallskip}
3867.4723 & $-$2.038 & NIST5 & 169086.766 & profile & $-$2.155   & \\
3867.4837 & $-$2.260 & NIST5 & 169086.843 & profile & $-$2.155   & \\
3867.6315 & $-$2.737 & NIST5 & 169087.631 & profile & $-$2.155   & \\
4009.2565 & $-$1.447 & NIST5 & 171134.897 & profile & $-$2.155   & \\   
4026.1844 & $-$2.628 & NIST5 & 169086.766 & profile & $-$2.155:: & no fit \\
4026.1859 & $-$1.453 & NIST5 & 169086.766 & profile & $-$2.155:: & \\
4026.1860 & $-$0.704 & NIST5 & 169086.766 & profile & $-$2.155:: & \\
4026.1968 & $-$1.453 & NIST5 & 169086.843 & profile & $-$2.155:: & \\
4026.1983 & $-$0.976 & NIST5 & 169086.843 & profile & $-$2.155:: & \\
\hline
\noalign{\smallskip}
\end{tabular}
\end{table*}

The most overabundant element is Xe ([$+$5.22]), followed by Hg ([$+$4.75]),
Ga ([$+$3.97]), P ([$+$2.24]), Mn ([$+$1.71]), Fe [($+$0.73)], and 
Ti ([$+$0.67]). 
The elements Ne, Si, Ca, and Cr are marginally overabundant. We note that a nearly solar abundance of 
$-$4.45 dex, rather than the average value of $-$4.37 dex derived from the equivalent widths, better 
reproduces with the synthetic spectrum most of the observed Si lines.

Helium is underabundant ($\sim$ [$-$1.1]), but it is 
difficult to state a definite abundance value, because some observed 
profiles cannot be fitted by the computed ones, whichever is the adopted
abundance. These are the lines at 4026, 4387, 4471, and 4921\,\AA{}.
We assumed an average abundance of N(\ion{He})/N$_{\rm tot}=0.007$, which 
reproduces rather well most of the lines listed in Table~\ref{tab:line_abundance}.
The wings of the lines at 4026, 4471, 5075, and  6678\,\AA{} are rather 
well reproduced by an average abundance of $-$2.17\,dex derived from 
all the \ion{He}{i} lines, but the observed  core is
weaker than the computed one.
This kind of behaviour, common to
several HgMn stars (e.g. \citealt{cast2004}), is ascribed to vertical abundance stratification.
The whole line at 4388\,\AA{}  
and the red wing of the line at 4922\,\AA{}  can not be fitted. 
The cause could be due to the several blends affecting them.
The other \ion{He}{i} lines at 3867, 4009, 4121, 4713, 5015, and 5047 are well
reproduced by the $-$2.17\,dex abundance. Some of them have minor contributions
of blends.   

Other underabundant elements are Al ($\le$[$-$1.2]), S ([$-$1.1]), 
O ([$-$0.69]), C ([$-$0.60]), and Mg ([$-$0.60]).
Finally, Ni is marginally underabundant. 

We could identify  a few observed but not predicted lines as \ion{As}{ii}.  
In fact,  no \ion{As}{ii} lines
are included in our line list owing to the lack of $\log\,gf$ values 
and excitation potentials for them. We used \ion{As}{ii}
wavelengths listed in the NIST database to identify  the
lines observed at 4494.30, 5497.727, 5558.09, 5651.32, and 6170.27\,\AA{}
as due to \ion{As}{ii}.
Arsenic was reportedly also present in the 
HgMn stars 46\,Aql \citep{sad2001} and HD\,71066 \citep{yuce2011}.

No lines of rare earth elements, as well as no lines of
\ion{Y}{ii}, \ion{Pt}{ii}, and \ion{Au}{ii} were observed.
 
For comparison, the abundances from \citet{alonso2003} 
are listed in the last column of Table~\ref{tab:abundance}. There is a 
large disagreement for almost all elements, except for silicon and iron.   
The differences in the abundances for iron and silicon 
amount to 0.05\,dex for \ion{Fe}{ii},  0.15\,dex for \ion{Fe}{iii}, and 
 0.06\,dex for \ion{Si}{ii}. They can be related  with both the different
ATLAS9 parameters and the microturbulent velocities adopted for the abundance analysis. 
The ATLAS9  parameters in this study are $\teff=13\,870$\,K, $\logg=3.8$, 
$[{\rm M}/{\rm H}]=0.0$, while those of \citet{alonso2003} are $\teff=13\,350$\,K, $\logg=3.76$, $[{\rm M}/{\rm H}]=0.5$;
the microturbulent velocities are $\xi=0.0$\,km\,sec$^{-1}$ and 1.2\,km\,sec$^{-1}$, respectively.
However, the differences for all the other elements are too large
to be only due to the different choices for the parameters.
Furthermore, \citet{alonso2003} derived
abundances for numerous elements that we did not observe at all in our spectra,
while we identified and derived abundances for \ion{Ne}{i}, \ion{Ga}{ii},
and \ion{Xe}{ii} that were not mentioned at all by \citet{alonso2003},
although these elements are present with numerous lines. 
 Unfortunately, they did
not publish the list of lines and equivalent widths they used, so that any
further comparison is not possible.

The ATLAS12 model was preferred to ATLAS9 for final 
abundance determination, to ensure consistency with the SYNTHE code in computing the line profiles; however 
abundance values derived using ATLAS9 and ATLAS12 differ by no more than 0.05\,dex.
The observed and synthetic spectra are presented on F.~Castelli's web 
page\footnote{http://wwwuser.oats.inaf.it/castelli/hd19400/hd19400.html}
together with the line-by-line identification.

\subsection{Emission lines}
\label{sect:emission_lines}

Similar to the spectral behaviour of a number of HgMn stars, the lines of multiplet 13 of \ion{Mn}{ii} 
($\lambda\lambda$ 6122-6132\,\AA) appear to be affected by emission. In fact, a very weak emission 
is observed for the blends  $\lambda\lambda$ 6125.861, 6126.225\,\AA, 
while a well observable strong absorption is predicted at these wavelengths. 
Furthermore, the blends at $\lambda\lambda$
6122.432, 6122.807\,\AA, at  6128.726, 6129.019, 6129.237\,\AA, and
at 6130.793, 6131.011, 6131.918\,\AA\ are observed much weaker than computed,
so that the core could have been filled by emission.

Other CP stars showing this kind of emission are, for example,  3 Centauri A
\citep{sig2000}, \citep{wahl2004}, 
46 Aquilae \citep{sig2000}, 
HR\,6000 \citep{cast2007}, and HD\,71066 \citep{yuce2011}.
This phenomenon was explained either in the context of non-LTE line formation
\citep{sig2001} or as due to a possible fluorescence mechanism 
\citep{wahl2000}.

\subsection{Anomalous line profile shapes}

\begin{figure}
\centering
\includegraphics[width=0.34\textwidth]{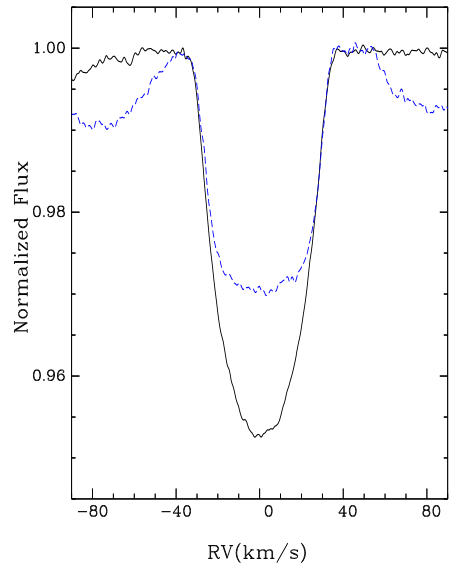}
\caption{Comparison of the average \ion{Mn}{ii} line profile (dashed line) with the 
average \ion{Fe}{ii} line profile (solid line).
}
\label{fig:mn}
\end{figure}

In all HARPS spectra the lines of \ion{Mn}{ii} and \ion{Hg}{ii}  present anomalous flat-bottom line profiles 
reminiscent of profile shapes observed in numerous HgMn stars (e.g., 
\citealt{Hubrig2006a,Hubrig2011}; \citealt{Mak2011a}), while the lines of other elements exhibit 
typical rotationally broadened line profiles. 
As an example, we display in Fig.~\ref{fig:mn} the average \ion{Mn}{ii} profile
overplotted with the average \ion{Fe}{ii} profile, using the best almost blend-free lines of moderate strength. 
To construct the 
average \ion{Mn}{ii} profile,
we employed the \ion{Mn}{ii} lines $\lambda\lambda$4292.2, 4363.3, 4478.6, 4518.9, 4738.3, 4755.7, and 4764.7. 
Three of them,  $\lambda\lambda$4738.3, 4755.7, and 4764.7,
have not been used in the abundance analysis because of their unknown hyperfine structure.
We note that two more \ion{Mn}{ii} lines,  $\lambda\lambda$4206.4 and 4365.2, were employed in the abundance analysis 
(see Table~\ref{tab:line_abundance}).
They are not included in our sample of lines selected for the search of variability, as the line
at $\lambda\lambda$4206.4 is slightly disturbed by a blend in the blue wing and the weak line
at $\lambda\lambda$4365.2 is considerably affected by noise at the third epoch.
In the calculation of the average \ion{Fe}{ii} profile, we used the \ion{Fe}{ii} lines $\lambda\lambda$4122.7, 4296.6, 4491.4, 4522.6, 
4923.9, 5002.0, and 5061.7, which constitute a subset of the sample of \ion{Fe}{ii} lines employed in the 
Fe abundance determination. 

\begin{figure}
\centering
\includegraphics[angle=90,width=0.42\textwidth]{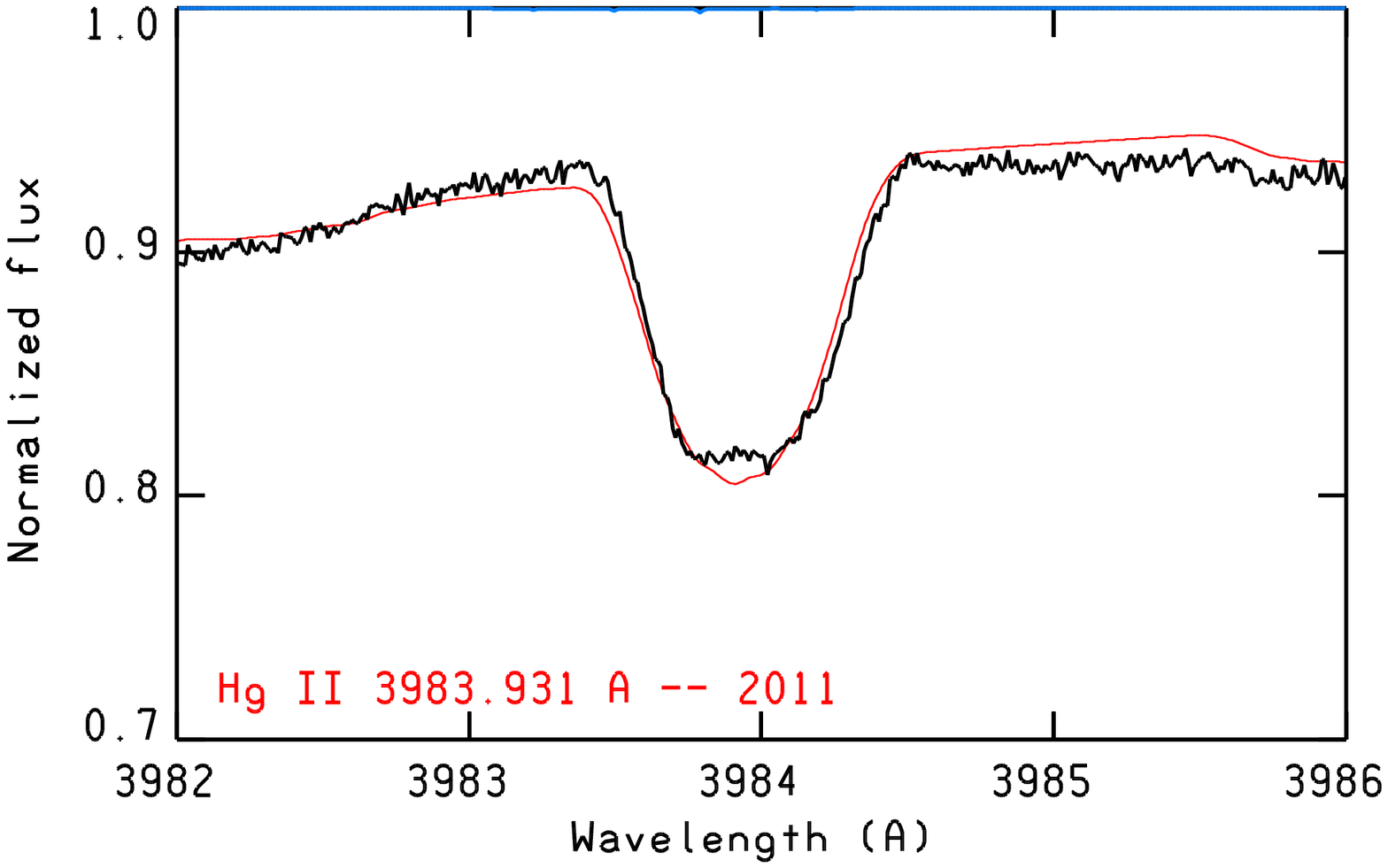}
\includegraphics[angle=90,width=0.42\textwidth]{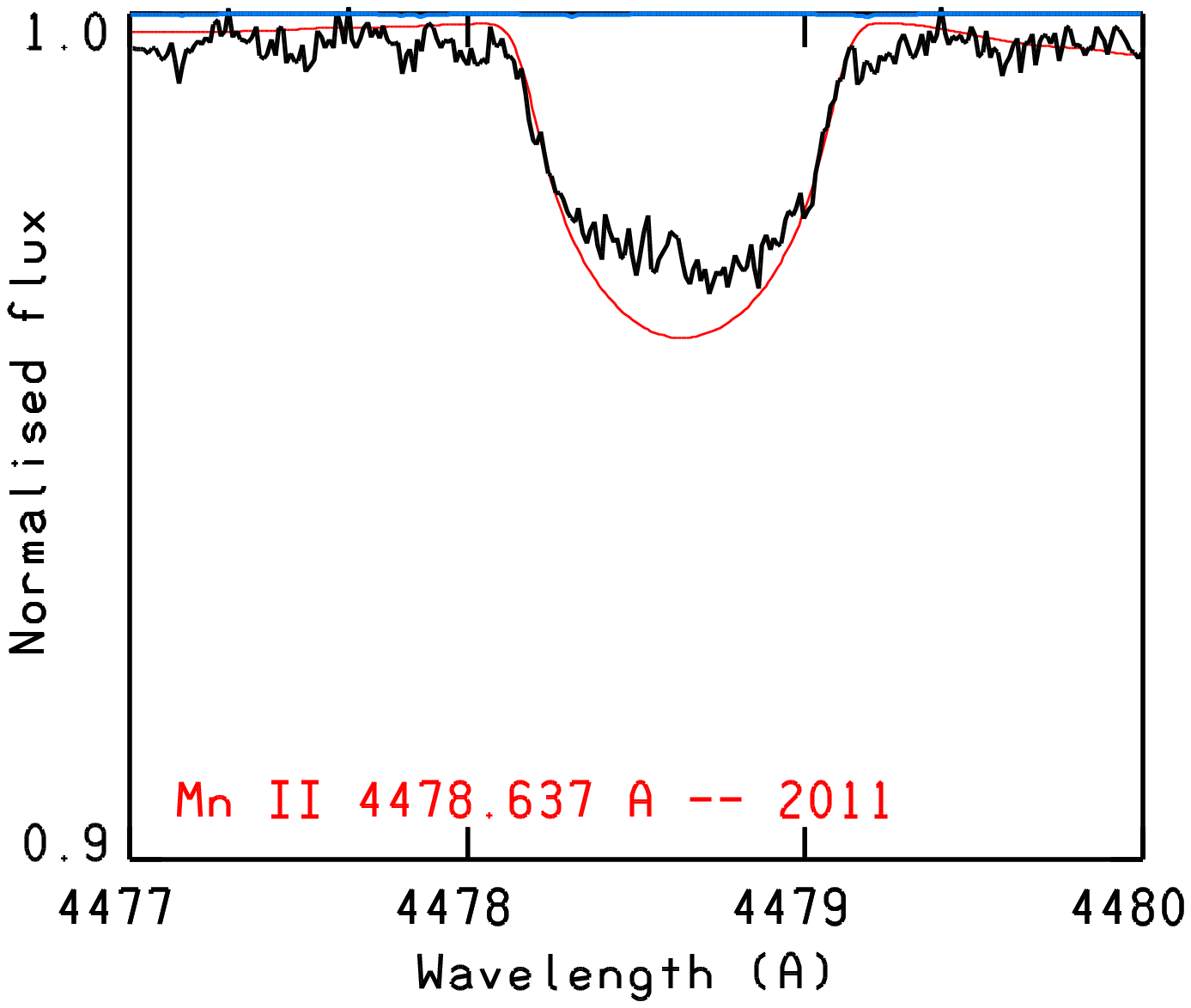}
\caption{
\ion{Hg}{ii} and \ion{Mn}{ii} line profiles observed in the HARPSpol 
spectrum obtained in 2011 highlighted in black together with
the synthetic profile shown by the thin red line.
The shape of the line profiles belonging to  \ion{Hg}{ii} and  \ion{Mn}{ii} deviates from the 
purely rotationally broadened profiles observed in the \ion{Fe}{ii}  and \ion{Cr}{ii} lines, 
indicating an inhomogeneous distribution of Hg and Mn on the stellar surface.
}
\label{fig:isot}
\end{figure}

The calculation of the synthetic spectrum for individual \ion{Mn}{ii} and \ion{Hg}{ii} lines 
indicates that the 
anomalous flat-bottom line profile shape is not caused by the presence of isotopic/hyperfine structure.
In the top panel of Fig.~\ref{fig:isot}, we display the synthetic profile of the \ion{Hg}{ii} $\lambda$3984 
line overplotted with the 
observed line profile. The bottom panel presents the synthetic and observed profiles of the 
\ion{Mn}{ii} $\lambda$4478 line.
The observed anomalous profile shape of both lines is reminiscent of the behaviour of line profiles 
of various elements in typical HgMn stars (e.g. \citealt{Hubrig2006a}). In spectroscopic binaries
these elements are frequently 
concentrated in non-uniform equatorial bands, which disappear exactly on the surface area, which
is permanently facing the secondary (e.g. \citealt{Hubrig2010}).

\section{Spectral variability}
\label{sect:variability}

For the study of the spectral variability we have on our disposal two HARPS spectra taken on two consecutive 
nights in 2011 December, while the third HARPS spectrum was obtained in 2013 July.
The spectra have different quality with a S/N of about 800 for observations in 2011
and a S/N of only about 500 for the observation in 2013.
To better understand the chemical spot pattern on the surface of HD\,19400, we decided to analyse the spectral variability 
by comparison of observations
separated by two timescales. On the one hand we compared the spectra from 2011 with each other to study the
day-to-day variations, and on the other hand, we compared the spectrum obtained in 2013 against 
the average of the two spectra from 2011 to search for  long term variation in the line profiles.

\begin{figure}
\centering
\includegraphics[angle=270,width=0.45\textwidth]{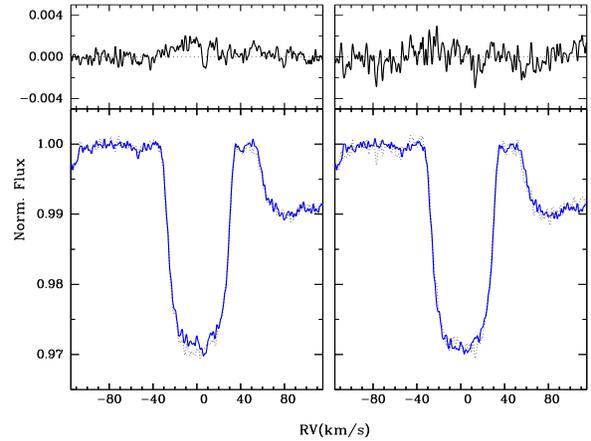}
\caption{
Profile variations of the Mn lines.
Left panel: day-to-day variations;  solid and dotted lines correspond to 2011 December 15 and 16,
respectively.
Right panel: year-to-year variations; solid and dotted lines present
2011 and 2013 observations, respectively.
The upper part of each panel shows the difference between the two plotted spectra.
}
\label{fig:varMn}
\end{figure}

In the left panel of Fig.~\ref{fig:varMn}, we compare the mean \ion{Mn}{ii} profiles obtained on two nights in 2011.
In the difference spectrum the rms of the noise in the nearby continuum is 
about 0.05\% while the variations are about 0.15\%.
The variability of the Mn lines is detected therefore at a level of 3$\sigma$.
We note that if the high-frequency noise is filtered, it becomes evident that the observed 
day-to-day variation is in fact four times larger than the $\sigma$ of the noise of similar
frequency.
The year-to-year variations are apparently not larger than the day-to-day variations,
and,  since the S/N of the spectrum obtained in 2013 is lower, the variations between 2011 and 2013
are not so clear. The flux differences, however, are still present at a level of about 2$\sigma$.

\begin{figure}
\centering
\includegraphics[angle=270,width=0.45\textwidth]{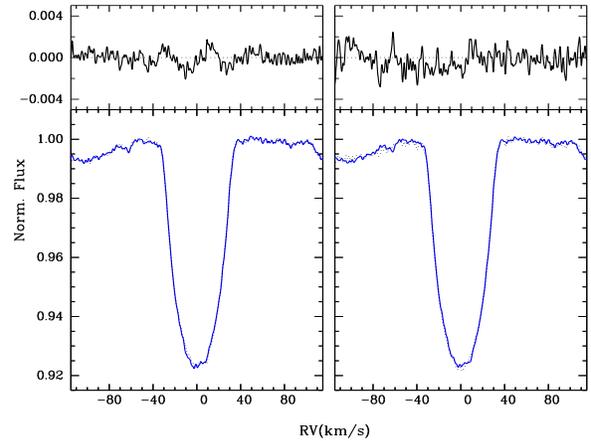}
\caption{
Profile variations of the Fe lines.
Left panel: day-to-day
variations;  solid and dotted lines correspond to 2011 December 15 and 16,
respectively.
Right panel: year-to-year variations; solid and dotted present
2011 and 2013 observations, respectively.
The upper part of each panel shows the difference between the two plotted spectra.
}
\label{fig:varFe}
\end{figure}

The same procedure as for the study of the Mn lines was applied to the Fe lines.
Figure~\ref{fig:varFe} shows the mean profile of the seven Fe lines. 
The day-to-day flux variations within the line profile are on the order of 0.13\%,
which is 3 times larger than the noise and of the order of 4$\sigma$, if high frequencies are filtered. 
Year-to-year flux differences are similar in size, but in this case represent
only a level of about 1.5 times over the noise and $\sim$2$\sigma$, after filtering high frequencies.

\begin{figure}
\centering
\includegraphics[angle=270,width=0.45\textwidth]{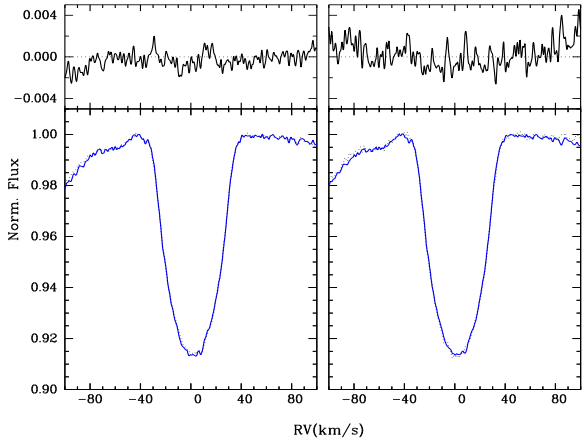}
\caption{
Profile variations of the P lines.
Left panel: day-to-day variations;  solid and dotted lines correspond to 2011 December 15 and 16,
respectively.
Right panel: year-to-year variations; solid and dotted lines present
2011 and 2013 observations, respectively.
The upper part of each panel shows the difference between the two plotted spectra.
}
\label{fig:varP}
\end{figure}

To study the variability of the \ion{P}{ii} lines, we selected the following six blend-free \ion{P}{ii} lines:
$\lambda\lambda$4420.71, 4530.82, 4589.85, 5386.90, 6024.18, and 6043.08.
The line profile difference between the two spectra taken in 2011 resembles that of the Fe lines
(Fig.\,\ref{fig:varP}).
However, the noise in this case is higher and the detection would be only at a level of 2$\sigma$.
The year-to-year comparison shows no significant variation, probably due to the rather high noise level.

\begin{figure}
\centering
\includegraphics[angle=270,width=0.45\textwidth]{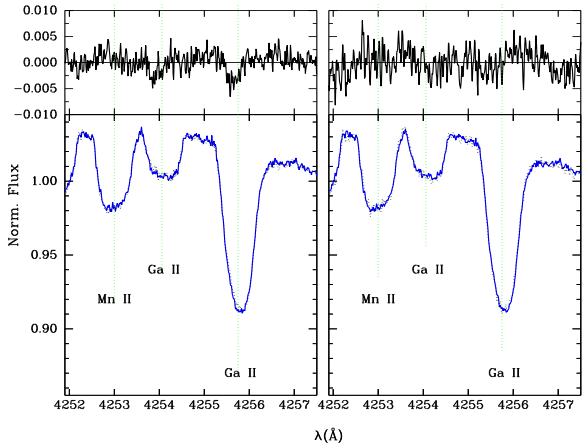}
\caption{
Profile variations of the Ga lines.
Left panel: day-to-day
variations;  solid and dotted lines correspond to 2011 December 15 and 16,
respectively.
Right panel: year-to-year variations;  solid and dotted lines present
2011 and 2013 observations, respectively.
The upper part of each panel shows the difference between the two plotted spectra.
}
\label{fig:varGa}
\end{figure}

The behaviour of line profiles belonging to other elements with strong overabundances, in particular Hg and Ga, 
also indicates a non-uniform
distribution. Unfortunately, for these elements, the number of useful lines is low and 
consequently the detection threshold is higher.
Figure~\ref{fig:varGa} shows the differences between the spectra around the only two clean
Ga lines: $\lambda$4254.0 and $\lambda$4255.7.
We note that the shape difference in the line profiles between spectra obtained on 2011 December 15 and 16
is the same for both Ga lines.
This supports the genuineness of the variation behaviour, even if the flux differences are not 
larger than 2$\sigma$. 
The differences with respect to the 2013 spectrum are within the noise level.

It is a remarkable finding that the \ion{Ga}{ii} lines appear indicative of variability.
Previous studies of He-weak magnetic Bp stars  with similar atmospheric parameters and with well established strong
magnetic fields also indicated strong overabundances of P, Ga, Xe, and a few heavy 
elements, such as Pt and Hg (e.g.\ \citealt{collado2009}).
Furthermore, a number of magnetic Bp stars with strongly overabundant
Ga display a large variation of Ga lines, which become the strongest at longitudinal magnetic 
field maxima, suggesting that Ga is accumulating near the magnetic poles (e.g.\ \citealt{artru1988}).
Interestingly, \citet{alecian1987} presented the radiative accelerations of gallium in Bp star atmospheres
and concluded that the presence of a magnetic field strongly modifies the gallium accumulation.

\begin{figure}
\centering
\includegraphics[angle=270,width=0.45\textwidth]{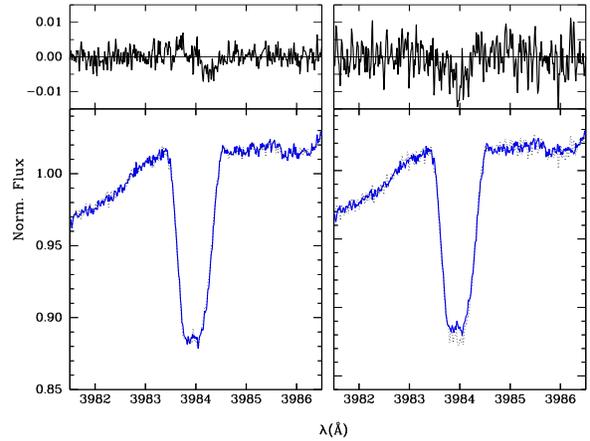}
\caption{
Profile variations of the Hg line.
Left panel: day-to-day variations;  solid and dotted lines correspond to 2011 December 15 and 16,
respectively.
Right panel: year-to-year variations; solid and dotted lines present
2011 and 2013 observations, respectively.
The upper part of each panel shows the difference between the two plotted spectra.
}
\label{fig:varHg}
\end{figure}

Also the Hg line at 3984\,\AA{} presents noticeable variations (Fig.~\ref{fig:varHg}).
Day-to-day variations are of the order of 0.4\%, while differences of about 0.8\%
are present between the 2011 and 2013 observations.
In both cases the variation is at a 2$\sigma$ level. If high-frequencies are filtered,
then the variations are at a 3$-$4$\sigma$ level.

Due to the small number of available spectra, it is currently not possible to decide whether the variability 
pattern of the line profiles belonging to different elements can be explained by an inhomogeneous element distribution 
on the stellar surface, or is due to pulsations.
Although no information on the rotational period is given in the literature, its upper limit can be estimated
from the measured $v\sin i$-value and the stellar radius. 
Using  $v\sin i=32.0\pm0.5$ km\,s$^{-1}$ and $R = 3.7\pm0.5$\,R$_{\odot}$ calculated using main-sequence evolutionary 
CL\'ES models \citep{Scu2008},
we obtain an upper limit of $P\le5.85\pm0.8$\,days. 
It is clear
that with such a rather low value for the rotation period, one can already expect to see weak variability
in observations obtained on two consecutive nights. 

\begin{figure}
\centering
\includegraphics[angle=270,width=0.42\textwidth]{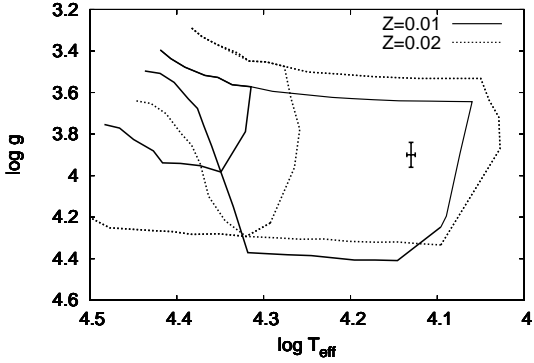}
\caption{
The position of the PGa star HD\,19400 in the H-R diagram.
The boundaries of the theoretical instability strips for $\beta$~Cep and SPB stars 
are taken from \citet{Miglio2007} for OP opacities. 
Full lines correspond to strips for metallicity $Z$ = 0.01 and dotted lines to strips with 
metallicity $Z$ = 0.02.
}
\label{fig:hr}
\end{figure}

In Fig.~\ref{fig:hr} we show the position of HD\,19400 in the log\,T$_{\rm eff}$--log\,$g$ diagram.  
together with the boundaries of the theoretical instability strips calculated for different metallicities 
($Z=0.01$ and $Z=0.02$) and using the OP opacities (http://cdsweb.u-strasbg.fr/topbase/op.html, 
see also \citealt{Miglio2007}).
The PGa star HD\,19400 with $\teff=13\,500$\,K and $\logg=3.90$ 
falls well inside the classical instability strip, where slowly pulsating B-type (SPB)
stars are found with expected pulsation periods from several hours to a few days.
Also classical magnetic Bp stars are located in the same region of the H-R diagram and frequently display 
strong overabundances of P, Ga, Xe, and heavy elements.

\begin{figure*}
\centering
\includegraphics[angle=270,width=0.495\textwidth]{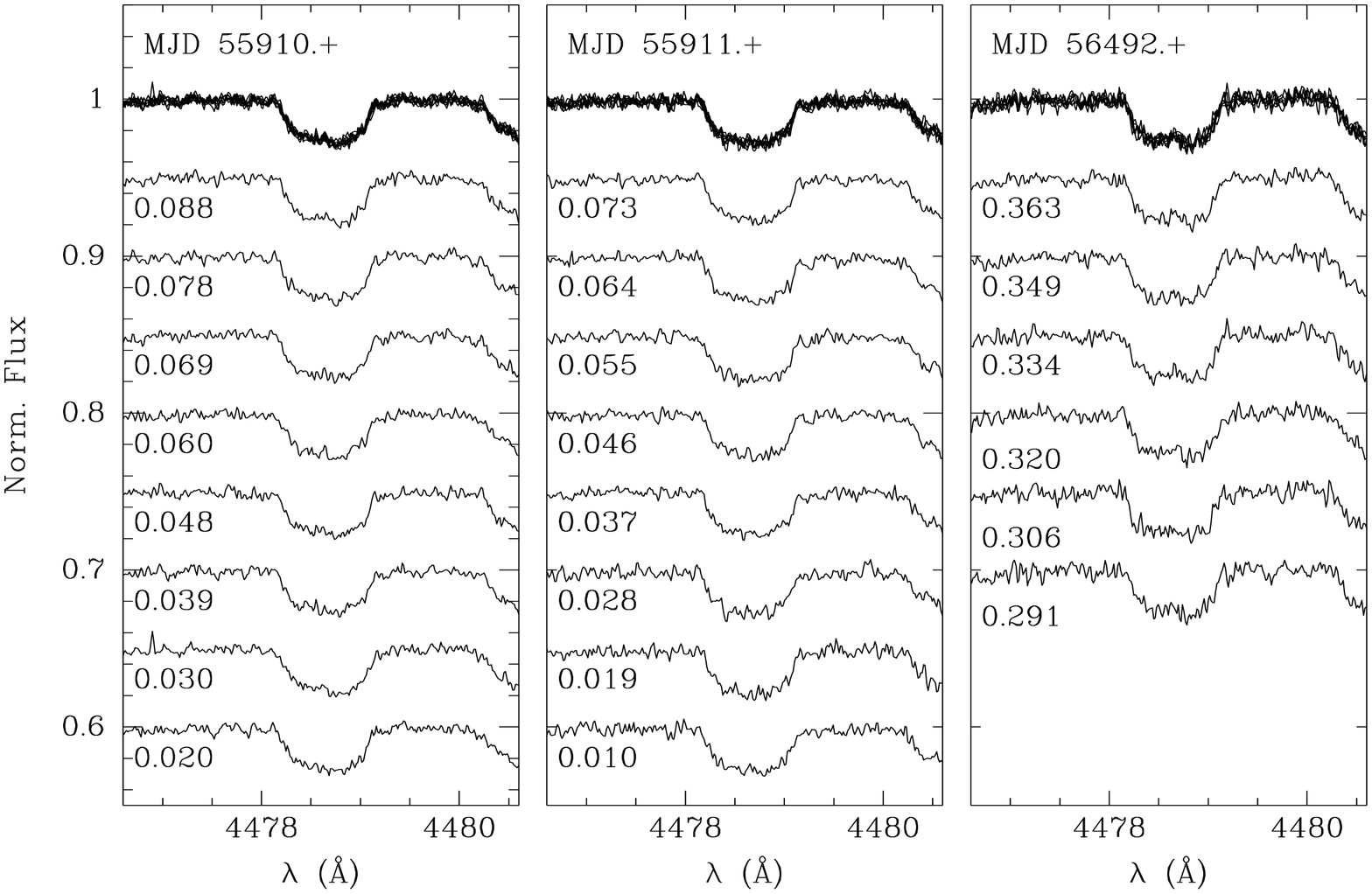}
\includegraphics[angle=270,width=0.495\textwidth]{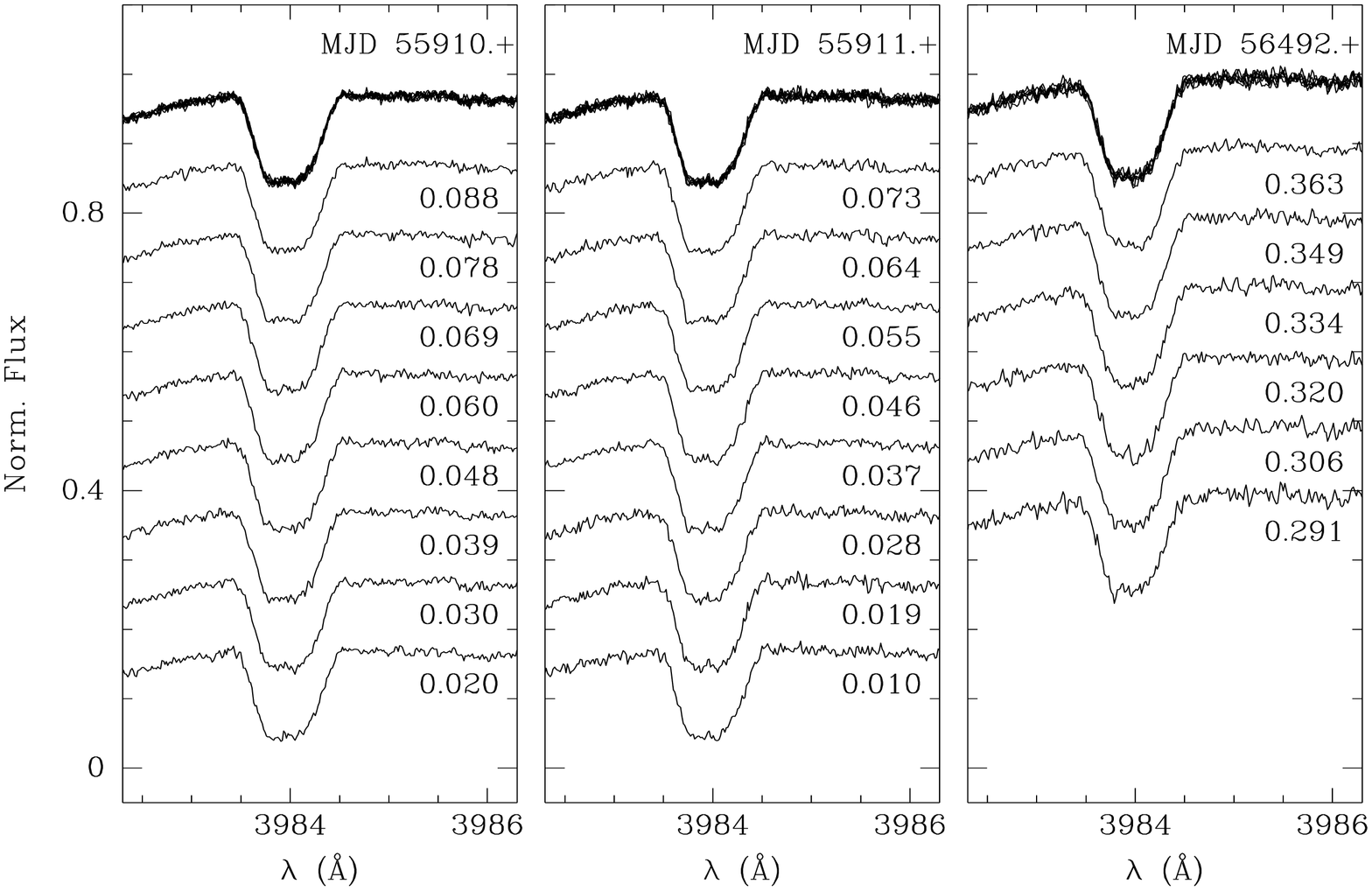}
\caption{
The behaviour of the line profiles of the  \ion{Mn}{ii} $\lambda$4478 line (left panel) 
and the \ion{Hg}{ii} $\lambda$3984 line (right panel) 
in HARPSpol subexposures obtained on 2011 December 15 and 16, and 
on 2013 July 19. The overplotted profiles are presented on the top.  
}
\label{fig:puls}
\end{figure*}

Since polarimetric observations usually consist of a number of short sub-exposures taken at 
different angles of the retarder wave plate, 
we used all available Stokes $I$ spectra for each sub-exposure to search for the presence of short-term variations in Mn and Hg line 
profiles. During both nights in 2011 December, the time difference between individual subexposures accounts for 13\,min,
while it is about 20\,min for the observations in 2013 July. 
In Fig.~\ref{fig:puls}, we present  the behaviour of line profiles of the \ion{Hg}{ii} $\lambda$3984 line, 
and  the \ion{Mn}{ii} $\lambda$4478 line, in HARPSpol subexposures obtained  on 2011 December 15 and 16 and 
on 2013 July 19.
No notable line profile variations above
the noise level, which is about 0.22\% of the continuum flux in the 2011 spectra
and about 0.32\% in 2013, are detected on these time scale.

\section{Magnetic field}
\label{sect:magn}

\begin{table}
\caption[]{
Previous magnetic field measurements of HD\,19400 using FORS\,1/2.
In the first column, we list the modified Julian date of mid-exposure followed by the measurements of the mean
longitudinal magnetic field $\left<B_{\rm z}\right>_{\rm all}$ using all available spectral lines
and $\left<B_{\rm z}\right>_{\rm hyd}$ using only
hydrogen lines. All quoted errors are 1$\sigma$ uncertainties.
}
\label{tab:log_measfors}
\centering
\begin{tabular}{cr@{$\pm$}lr@{$\pm$}l}
\hline \hline\\[-7pt]
\multicolumn{1}{c}{MJD} &
\multicolumn{2}{c}{$\left<B_{\rm z}\right>_{\rm all}$ [G]} &
\multicolumn{2}{c}{$\left<B_{\rm z}\right>_{\rm hyd}$ [G]} \\
\hline\\[-7pt]
  52852.371       &   151 & 46 &   217 & 65 \\
  55845.295       &    14 & 24 &    32 & 26 \\
  55935.109       & $-$65 & 26 & $-$110& 30 \\
\hline\\
\end{tabular}
\end{table}

A few polarimetric spectra of HD\,19400  were previously obtained with 
FORS\,1 \citep{Hubrig2006b},
and most recently with FORS\,2 on Antu (UT1) from 2011 May to 2012 January \citep{Hubrig2012} at the rather
low resolution of $\sim$2000.
The magnetic field measurements from these earlier data
are presented in Table~\ref{tab:log_measfors} together with the modified Julian date of mid-exposure. 
Out of the three measurements, weak magnetic field detections 
at a 3$\sigma$ significance level were achieved on two different epochs.
Given the low resolution  of FORS\,1/2, these spectra do not allow however
to measure the longitudinal magnetic field on lines of individual elements separately.

We note that the magnetic field topology in HgMn stars is currently unknown.
The recent study of \citet{Hubrig2012} seems to indicate the existence of intriguing correlations 
between the strength of the magnetic field, abundance anomalies, and binary properties.
Measurement results for a few stars revealed that element underabundance (respectively overabundance) is 
observed where the polarity of the magnetic field is negative (respectively positive).
An inhomogeneous chemical abundance distribution is observed  most frequently on 
the surface of upper-main sequence
Ap/Bp stars with large-scale organised magnetic fields.
The abundance distribution of certain elements in these stars is 
usually non-uniform and non-symmetric with respect to the rotation axis, but shows a kind of symmetry between the 
topology of the magnetic field and the element distribution. 
Assuming that a similar kind of symmetry exists in HgMn and PGa stars, 
it appears reasonable to use the \ion{Mn}{ii} lines for magnetic field measurements since \ion{Mn}{ii} 
shows the strongest variability in the
spectra of HD\,19400.

\begin{table}
\caption{Shifts between the line centres of
gravity in the right and left circularly polarised spectra obtained
on three different nights. In the first column, we list the wavelengths
of the \ion{Mn}{ii} lines followed by their corresponding Land\'e factors and 
the wavelength shifts in \AA{} on each epoch.  
}
\label{tab:results}
\begin{tabular}{rrrrr}
\hline
\hline\noalign{\smallskip}
\multicolumn{1}{c}{$\lambda$} &
\multicolumn{1}{c}{$g_{\rm eff}$} &
\multicolumn{3}{c}{$\Delta\lambda$ [\AA{}]} \\
\multicolumn{1}{c}{[\AA{}]} &
\multicolumn{1}{c}{} &
\multicolumn{1}{c}{Night 1} &
\multicolumn{1}{c}{Night 2} &
\multicolumn{1}{c}{Night 3} \\
\hline
4000.033 & 1.076 &            &     0.0001 &     0.0014 \\
4110.615 & 1.002 &  $-$0.0010 &     0.0004 &            \\
4184.454 & 1.065 &  $-$0.0024 &  $-$0.0005 &            \\
4206.367 & 1.362 &  $-$0.0001 &     0.0020 &  $-$0.0013 \\
4240.390 & 0.929 &  $-$0.0046 &     0.0040 &     0.0024 \\
4259.200 & 1.315 &  $-$0.0009 &  $-$0.0016 &  $-$0.0013 \\
4260.467 & 2.725 &  $-$0.0027 &     0.0040 &     0.0086 \\
4292.237 & 1.270 &  $-$0.0001 &            &     0.0051 \\
4326.639 & 1.379 &  $-$0.0013 &     0.0004 &     0.0011 \\
4363.255 & 1.112 &  $-$0.0019 &  $-$0.0005 &     0.0043 \\  
4365.217 & 1.500 &            &  $-$0.0042 &     0.0067 \\     
4478.637 & 1.500 &  $-$0.0104 &     0.0026 &            \\
4518.956 & 1.507 &  $-$0.0007 &  $-$0.0013 &            \\
4727.841 & 0.506 &            &     0.0031 &     0.0029 \\
4730.395 & 0.712 &     0.0022 &  $-$0.0022 &            \\
4734.136 & 0.724 &  $-$0.0010 &            &            \\
4738.290 & 1.080 &     0.0005 &            &            \\
4755.727 & 1.058 &  $-$0.0029 &     0.0009 &            \\
4764.728 & 1.027 &  $-$0.0009 &     0.0019 &     0.0032 \\
4791.782 & 1.085 &  $-$0.0076 &  $-$0.0042 &  $-$0.0030 \\
4806.823 & 1.048 &     0.0013 &     0.0028 &            \\
4839.737 & 1.358 &  $-$0.0031 &  $-$0.0027 &     0.0090 \\
\hline
\end{tabular}
\end{table}

\begin{figure*}
\centering
\includegraphics[width=0.28\textwidth]{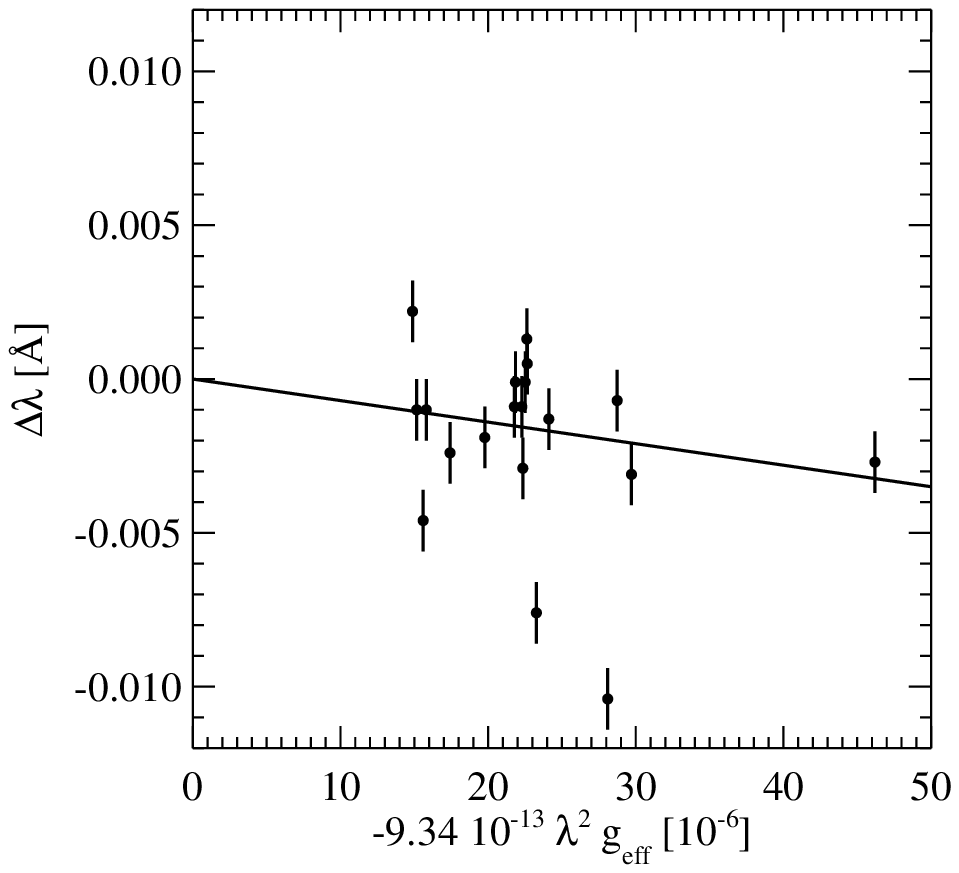}
\includegraphics[width=0.28\textwidth]{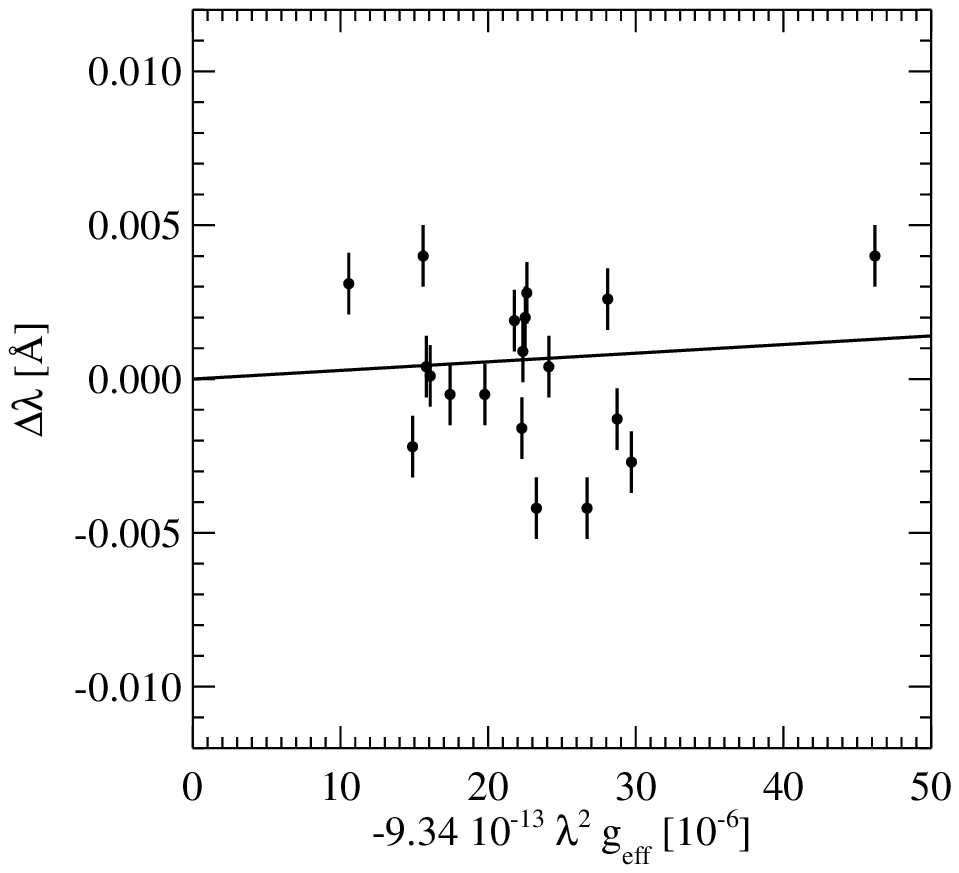}
\includegraphics[width=0.28\textwidth]{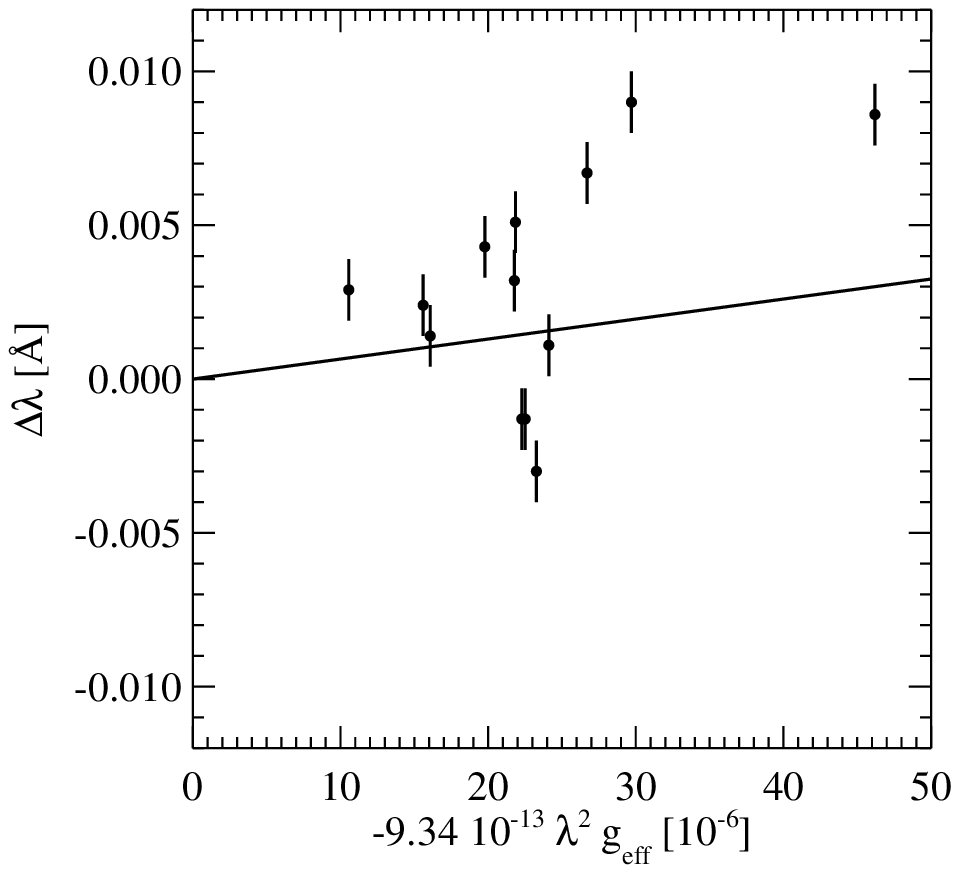}
\caption{
Linear regression analysis applied to the observations on three different nights.
For each line, the shift between the line centres of
gravity in the right and left circularly polarised spectra is plotted against 
$-9.34~10^{-13} \lambda^2 g_{\rm eff}$. The straight lines
represent the best fit resulting from a linear regression analysis.
}
\label{fig:regr}
\end{figure*}

The major problem in the analysis of high-resolution spectra is the proper line identification
of blend free spectral lines. The quality of the selection varies strongly from star to star,
depending on binarity, line broadening, and the richness of the spectrum. 
The best 22, mostly blend-free, \ion{Mn}{ii} lines, including also six \ion{Mn}{ii} lines used 
in the abundance determination, we employed in the diagnosis
of the magnetic field on the surface of HD\,19400 are presented in Table~\ref{tab:results}
together with their Land\'e factors.
The Land\'e factors were 
taken from Kurucz's list of atomic data\footnote{http://kurucz.cfa.harvard.edu/atoms}.
As a first step, we used for the measurements the moment technique developed by Mathys (e.g.\ \citealt{Mathys1991}).
This technique allows us not only the determination of the mean longitudinal magnetic field, but 
also to prove the presence of crossover effect and quadratic magnetic fields.
For each line, the measured shifts between the line profiles in the left- and right-hand circularly polarised HARPS spectra
are presented in Table~\ref{tab:results}.
The linear regression analysis in the
$\Delta\lambda$ versus $\lambda^2 g_{\rm eff}$ diagram, following the formalism discussed by 
\citet{Mathys1991,Mathys1994}, 
yields
values for the mean longitudinal magnetic field $\left< B_{\rm z} \right>$  between $-$70\,G and $+$65\,G.
A weak negative longitudinal magnetic field 
$\left<B_z\right>=-70\pm23$\,G at 3$\sigma$ level is measured on the first epoch, and  
$\left<B_z\right>=65\pm30$\,G at 2.2$\sigma$ level on the third epoch. 

\begin{table}
\caption[]{
Magnetic field measurements of HD\,19400 using HARPS.
In the first column, we list the modified Julian date of mid-exposure followed by the measurements of the mean
longitudinal magnetic field using \ion{Mn}{ii} lines from polarised spectra and null polarisation spectra.
All quoted errors are 1$\sigma$ uncertainties.
}
\label{tab:log_meas}
\centering
\begin{tabular}{cr@{$\pm$}lr@{$\pm$}l}
\hline \hline\\[-7pt]
\multicolumn{1}{c}{MJD} &
\multicolumn{2}{c}{$\left<B_{\rm z}\right>_{\rm Mn}$ [G]} &
\multicolumn{2}{c}{$\left<B_{\rm z}\right>_{\rm Mn,N}$ [G]} \\
\hline\\[-7pt]
  55910.054       &  $-$70 & 23 & $-$13 & 24\\
  55911.042       &    28 & 18 &  $-$18 & 21 \\
  56492.327       &  65 & 30 &  25  & 32 \\
\hline\\
\end{tabular}
\end{table}

Further, the mean longitudinal magnetic fields $\left<B_{\rm z}\right>_{\rm Mn,N}$
were measured from null polarisation spectra, which are calculated by combining the subexposures 
in such a way that polarisation cancels out. The results of our measurements are presented in Table~\ref{tab:log_meas}, 
and the corresponding linear regression plots are shown in Fig.~\ref{fig:regr}.
The measurements on the spectral lines of \ion{Mn}{ii} 
using null spectra are labeled by $N$ in Table~\ref{tab:log_meas}.
Since no significant fields could be determined from null spectra, we conclude that
any noticeable spurious polarisation is absent. No significant crossover and mean quadratic magnetic field have been 
detected on the three observing epochs. 

Our assumption that the inhomogeneous distribution of  \ion{Mn}{ii} over the stellar surface
is due to the action of a magnetic field does not exclude the possibility that also ions with less pronounced line
profile variability are inhomogeneously distributed across the stellar surface. Indeed, they could be concentrated towards 
the rotation poles, or have a predominantly symmetric distribution about the rotation axis.
Among the elements showing less pronounced line
profile variations in the spectra, the most numerous lines belong to \ion{Fe}{ii} and \ion{P}{ii}.
Additional magnetic field measurements have been thus carried out using 35 \ion{Fe}{ii} and 33 \ion{P}{ii} lines
listed in Table~\ref{tab:line_abundance}.
Neither measurements using the \ion{Fe}{ii} lines nor the \ion{P}{ii} lines showed evidence
for the presence of a magnetic field.

During the last few years, a number of attempts to detect mean longitudinal magnetic fields in HgMn stars have been made 
by several authors using the line addition technique, the least-squares deconvolution (LSD;
e.g.\ \citealt{Mak2011a,Mak2011b}), and the multi-line Singular Value Decomposition (SVD) technique 
\citep{Hubrig2014}. A high level of precision, from a few to tens of Gauss, 
is achieved through application of these techniques \citep{Donati1997,carroll2012}, 
which combine hundreds of spectral lines 
of various elements. In such techniques an assumption is made that all spectral lines are identical in shape and can be described 
by a scaled mean profile. However, the lines of different elements with different abundance distributions across 
the stellar surface sample the magnetic field in different manners. Combining them, as is done with such techniques,
may lead to the dilution of the magnetic signal or even to its (partial) cancellation, if enhancements of different 
elements occur in regions of opposite magnetic polarities. 
A shortcoming of both techniques is that a high level of precision is achievable only if a large number
of lines is involved in the analysis.   

\begin{figure*}
\centering
\includegraphics[width=0.32\textwidth]{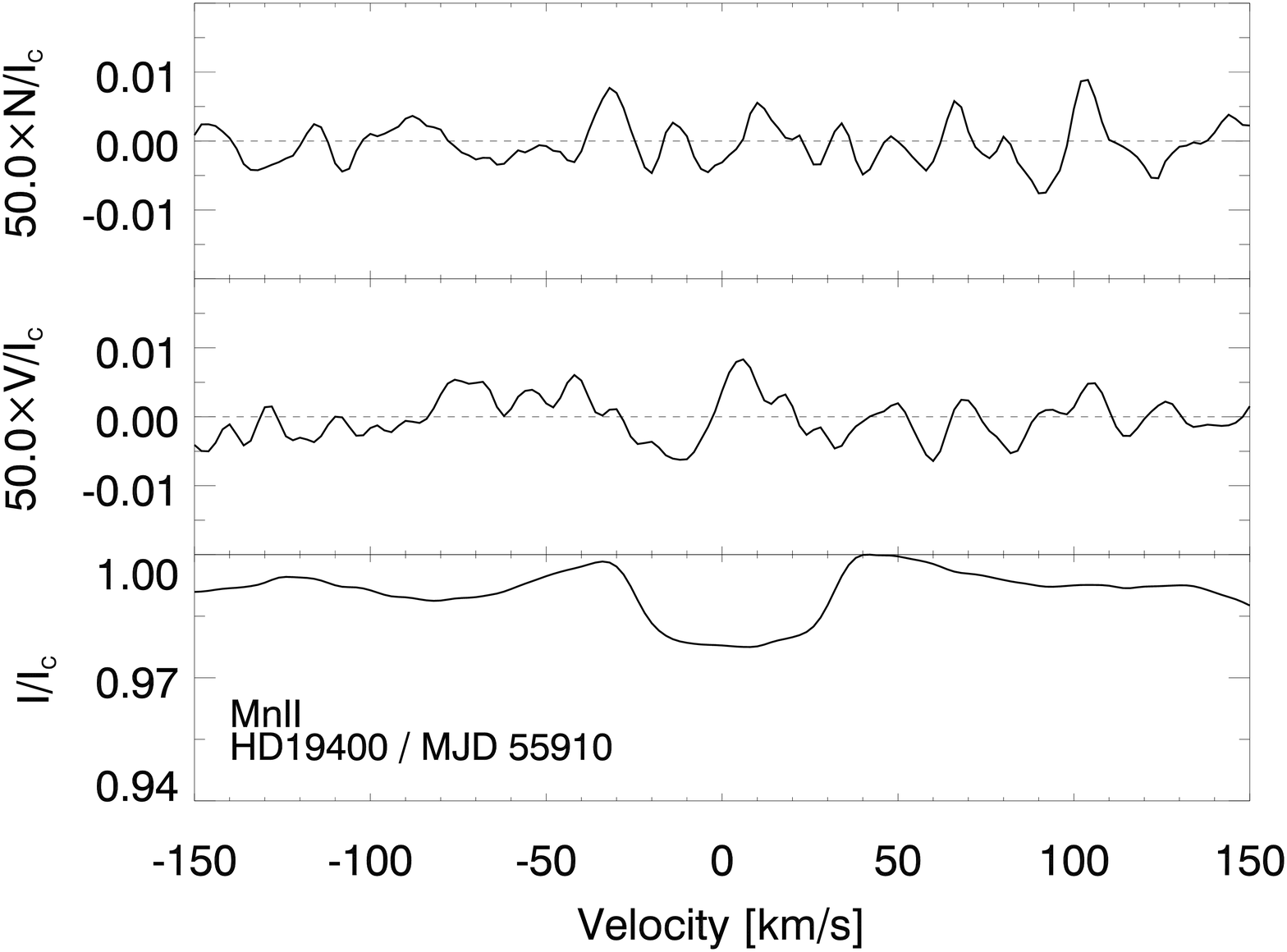}
\includegraphics[width=0.32\textwidth]{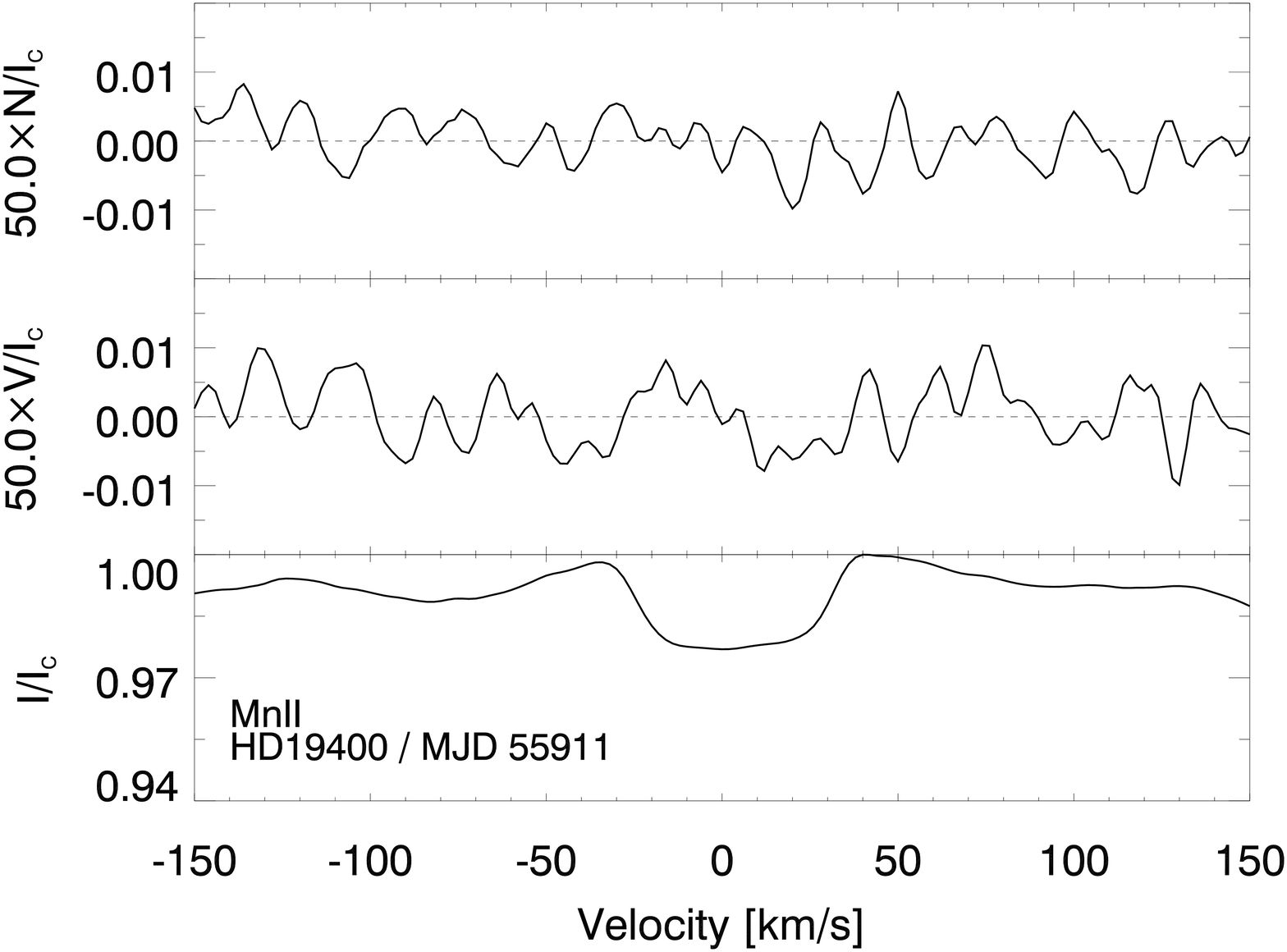}
\includegraphics[width=0.32\textwidth]{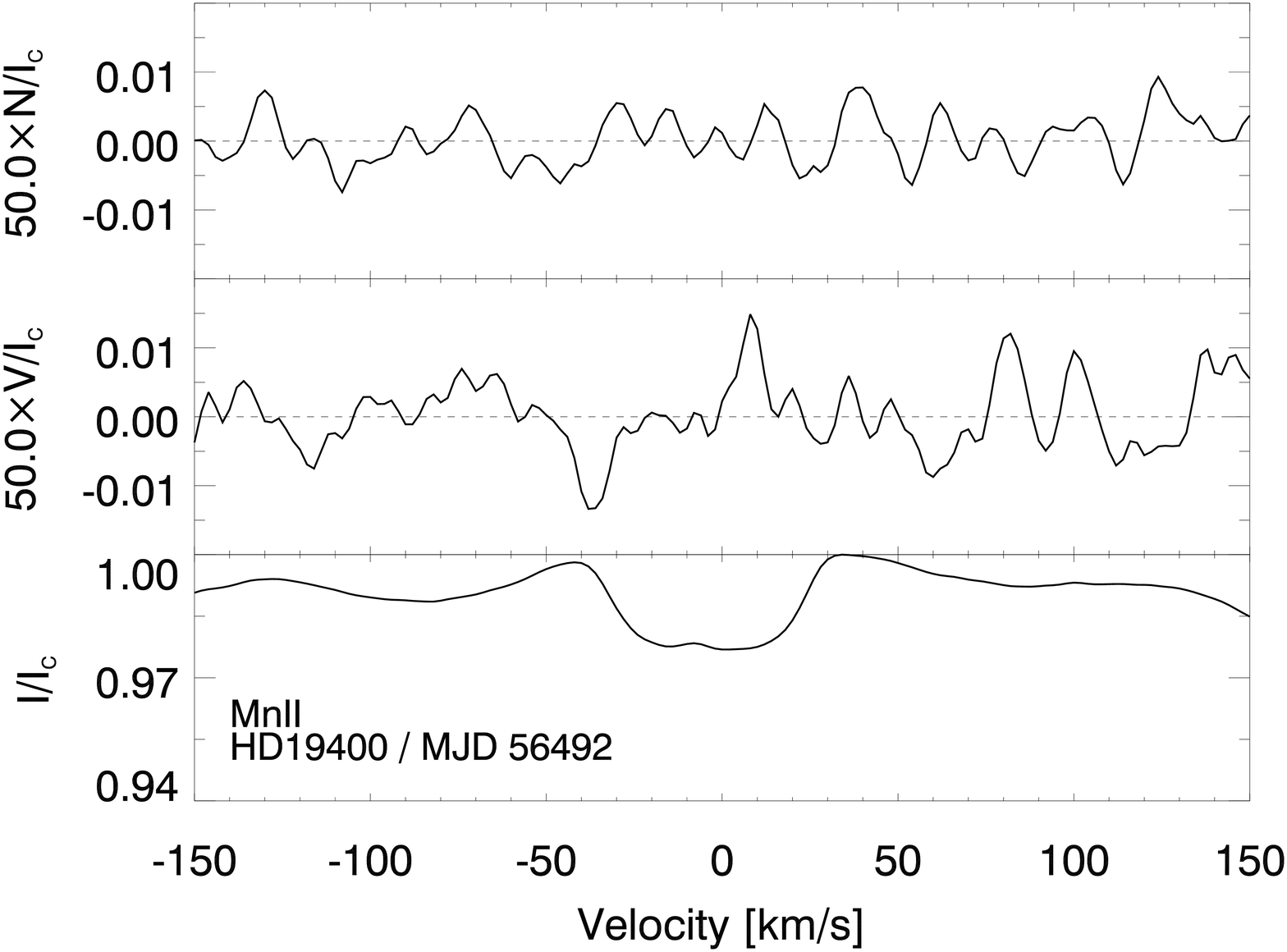}
\caption{
From left to right, we present the \ion{Mn}{ii} SVD profiles of HD\,19400 obtained at
the three different epochs from polarised spectra and null polarisation spectra.
From bottom to top, one can see the $I$, $V$, and $N$ profiles.
The $V$ and $N$ profiles were expanded by a factor of 50 for better visibility.
}
\label{fig:SVD1}
\end{figure*}

In Fig.~\ref{fig:SVD1}, we present the results of the SVD analysis of the observations on all three epochs
using 32~\ion{Mn}{ii} 
lines selected from the VALD database (e.g.,\ \citealt{pisk1995}; \citealt{kupka2000}).
We note that although the S/N of the two datasets obtained in 2011 is higher than for the most recent observation,
the reconstruction for the last observation was
performed with a smaller number of Eigenprofiles,
which results in a smaller relative noise contribution.
As the noise level scales with the effective
dimension of the signal subspace (see Sect.~3 of \citealt{carroll2012}),
the noise levels of the reconstructed profiles for all three nights are approximately the same.
On the first epoch, the measurement using the SVD technique shows the longitudinal
magnetic field $\left<B_z\right>=-76\pm25$\,G.
Using the false alarm probability (FAP; \citealt{Donati1992}) in the region of the 
whole Stokes $I$ line profile, we obtain for this measurement $FAP=0.008$.
According to \citet{Donati1992}, an
FAP smaller than 10$^{-5}$ can be considered as definite detection, while $10^{-5} < FAP < 10^{-3}$
are considered as marginal detections. 
We note that at this epoch the Zeeman feature is well visible
in the Stokes $V$ spectrum. However, the obtained FAP value is too high for 
a marginal detection.
The interesting fact is that the observed Stokes $V$ Zeeman 
feature is slightly shifted to the blue from the line center.

For the second epoch, we measure the mean longitudinal magnetic field $\left<B_z\right>=9\pm35$\,G with
the corresponding FAP value of 0.045. The observed Zeeman feature  in the SVD Stokes $V$ profile is 
reminiscent of a typical
crossover profile, and, similar to the Zeeman feature in the first epoch, is also slightly shifted to the 
blue from the line center.
The corresponding FAP values are 0.045 for the second epoch and  0.25 for the third epoch. 
Especially intriguing is the appearance of the SVD Stokes $I$ profile with the 
corresponding Zeeman feature  in the SVD Stokes $V$ profile obtained 
for the third epoch. The shape of the SVD Stokes $I$ profile shows a slight splitting shifted from the line center,
indicating that we likely observe in this phase two Mn surface spots.  
If we assume that the Zeeman feature in the SVD Stokes $V$ profile is not due to pure noise, then the observed negative and 
positive peaks 
in the Zeeman feature could probably correspond to two different Mn spots. The measured field for these features
indicates a magnetic field of about $-35$\,G for the negative peak and $+35$\,G for the positive peak with $FAP=0.029$.
On the other hand, it is clear that the amplitude of the features inside the SVD profiles is comparable to the amplitude of 
the noise outside the SVD profiles.

A similar analysis using the SVD technique was carried out for 115 \ion{Fe}{ii} and 33 \ion{P}{ii} lines selected from the VALD database.
The measurement on the first epoch using \ion{Fe}{ii} shows a longitudinal
magnetic field $\left<B_z\right>=-91\pm35$\,G with a false alarm probability
$FAP=0.003$. For the second epoch, we obtained $\left<B_z\right>=-28\pm21$\,G with $FAP=0.009$. 
No indications for a probable presence of a weak magnetic field was found on the third epoch.
The obtained FAP value for the measurement
on the first epoch is lower than that obtained for the measurement on the \ion{Mn}{ii} lines,
but is still three times too high for a marginal detection.

\begin{figure*}
\centering
\includegraphics[width=0.32\textwidth]{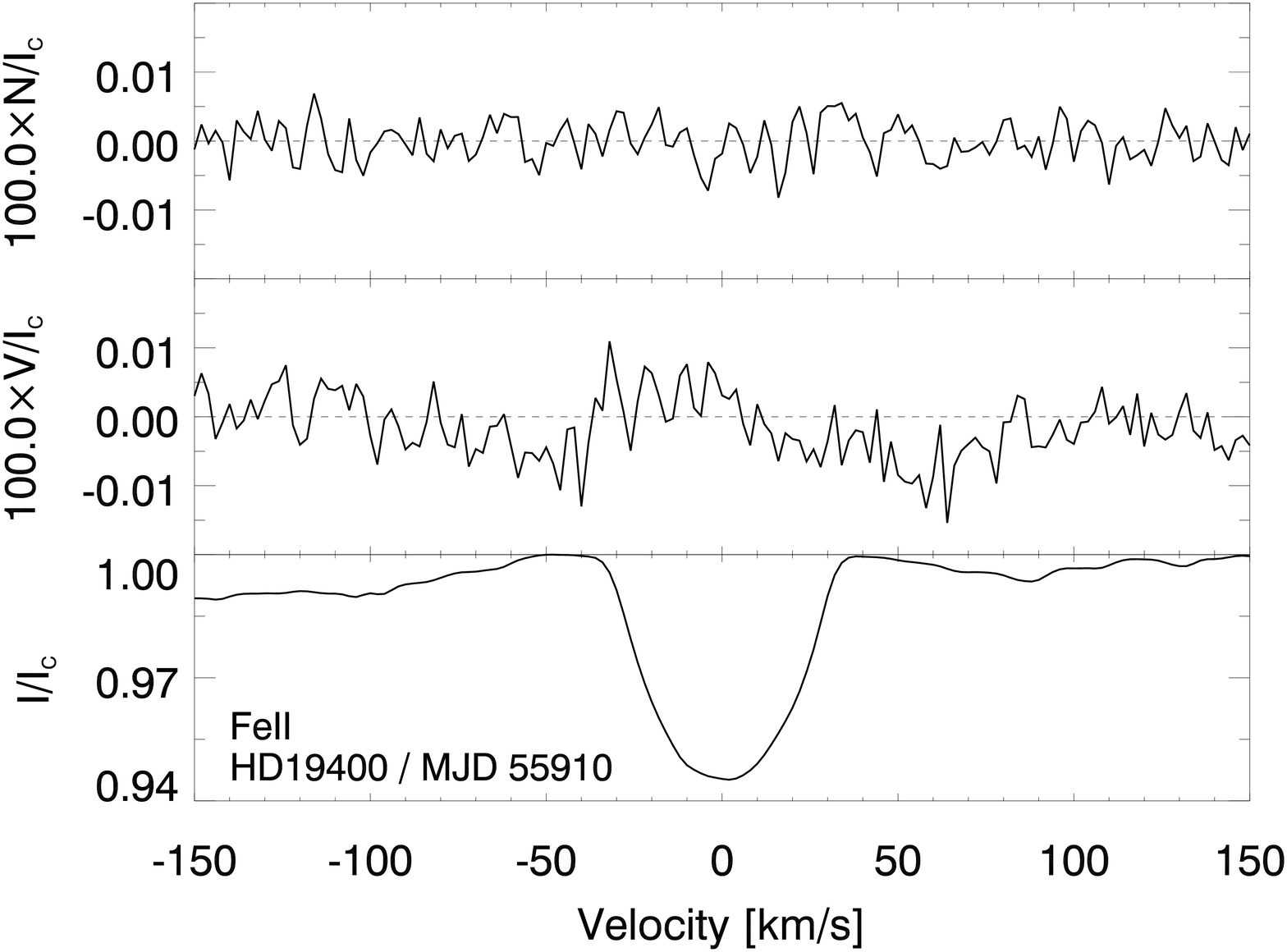}
\includegraphics[width=0.32\textwidth]{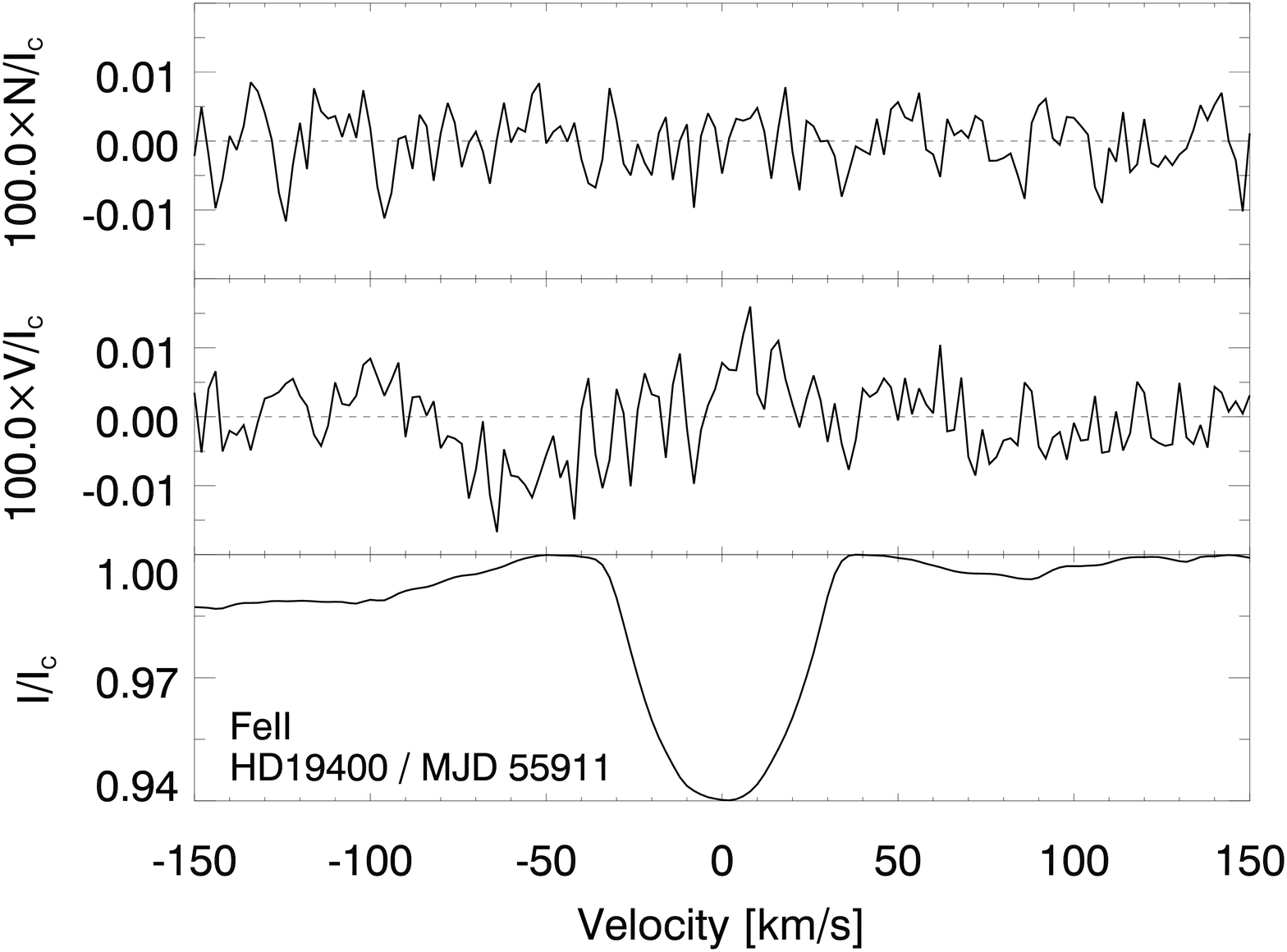}
\includegraphics[width=0.32\textwidth]{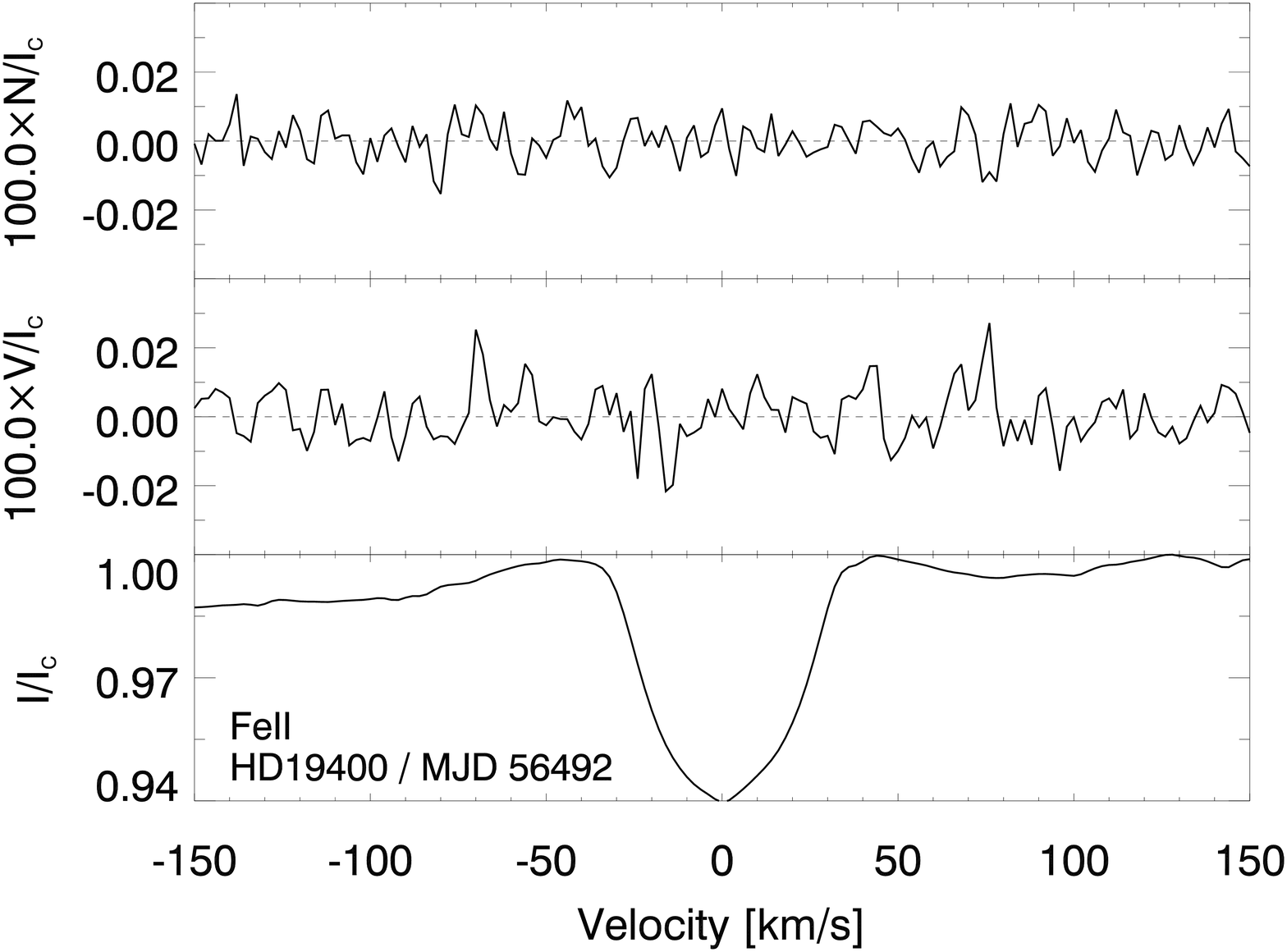}
\caption{
From left to right, we present the \ion{Fe}{ii} SVD profiles of HD\,19400 obtained at
the three different epochs from polarised spectra and null polarisation spectra.
From bottom to top, one can see the $I$, $V$, and $N$ profiles.
The $V$ and $N$ profiles were expanded by a factor of 100 for better visibility.
}
\label{fig:SVD1_Fe}
\end{figure*}

\begin{figure*}
\centering
\includegraphics[width=0.32\textwidth]{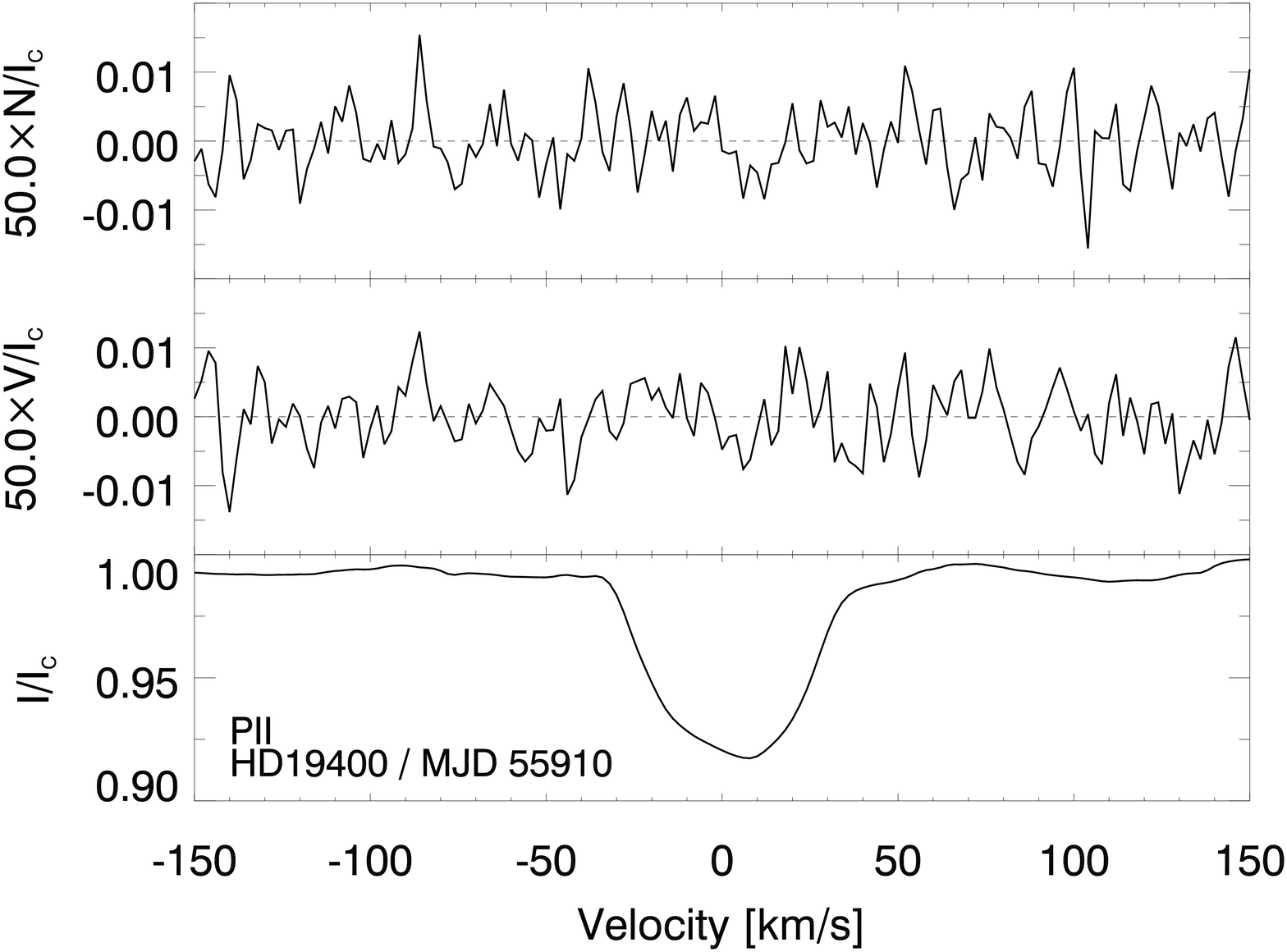}
\includegraphics[width=0.32\textwidth]{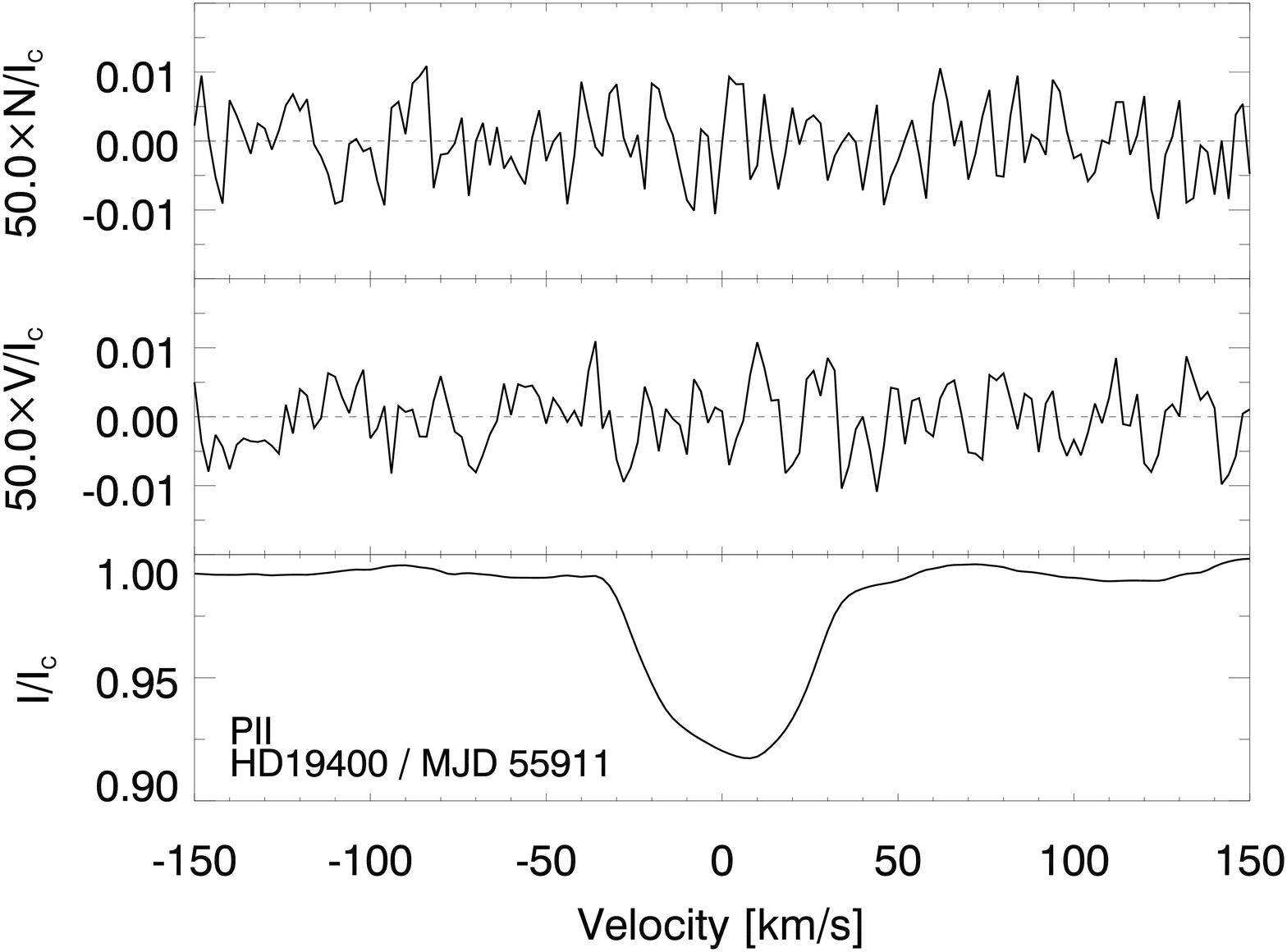}
\includegraphics[width=0.32\textwidth]{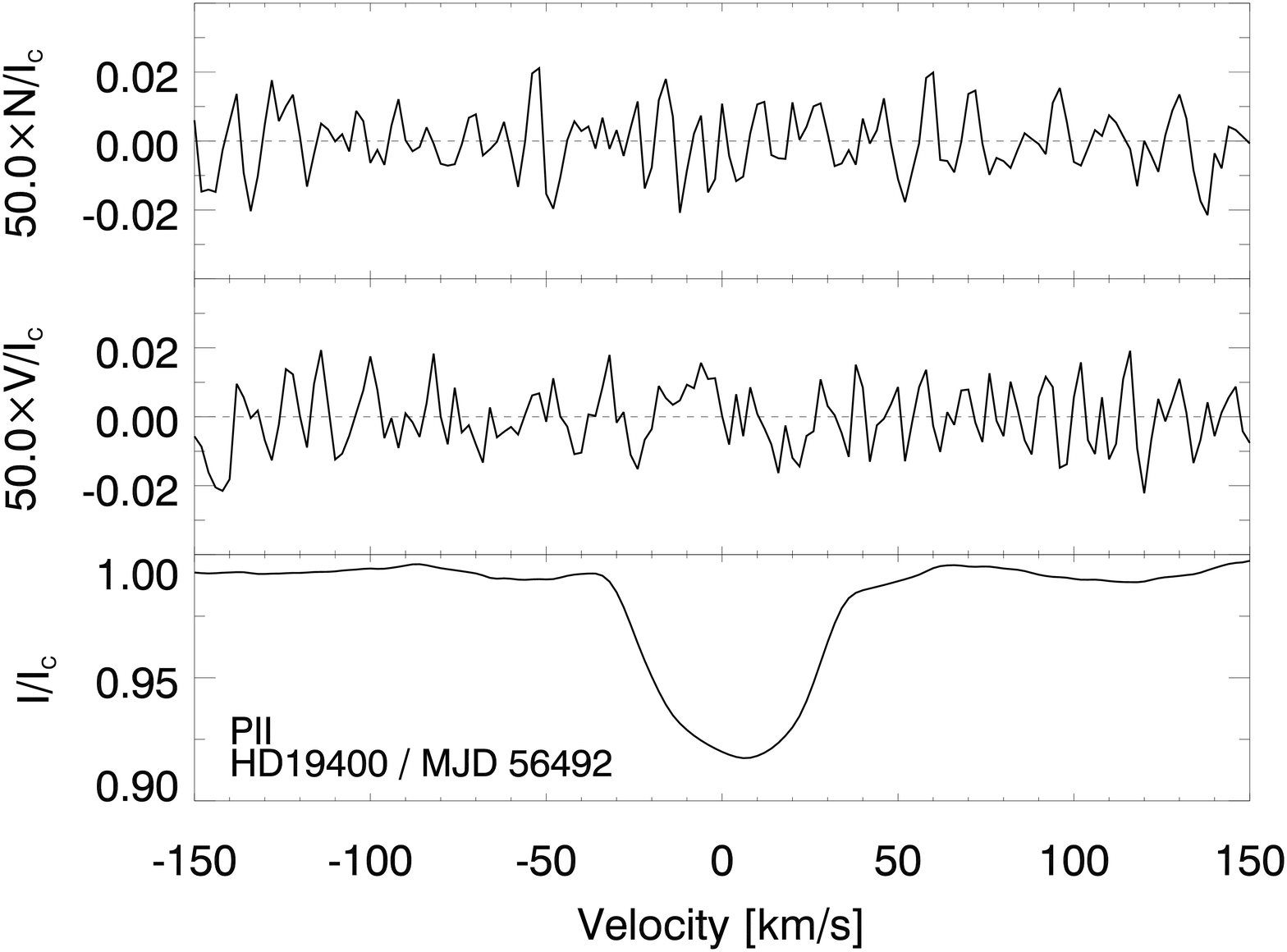}
\caption{
From left to right, we present the \ion{P}{ii} SVD profiles of HD\,19400 obtained at
the three different epochs from polarised spectra and null polarisation spectra.
From bottom to top, one can see the $I$, $V$, and $N$ profiles.
The $V$ and $N$ profiles were expanded by a factor of 50 for better visibility.
}
\label{fig:SVD1_P}
\end{figure*}

No field detection was achieved in the analysis using the \ion{P}{ii} lines, where 
the Stokes~$V$ spectra appear rather flat on all three epochs.
In Figs.~\ref{fig:SVD1_Fe} and \ref{fig:SVD1_P}, we present the results 
of the SVD analysis using the \ion{Fe}{ii} and \ion{P}{ii} lines, respectively.

In summary, although some indications on the probable existence of a weak magnetic field in HD\,19400  are found
in our analysis,
no definite conclusion on the presence of the magnetic field and its topology can currently be drawn.
Obviously, more high-resolution, high S/N polarimetric observations are urgently 
needed to properly understand the nature of this type of stars.

\section{Discussion}
\label{sect:disc}

In this work, we used high quality HARPSpol spectra of the PGa star HD\,19400 to carry out an abundance analysis, 
search for spectral variability, and the presence of a weak magnetic field.
We present the abundances of various elements determined using an ATLAS12 model, including the abundances of a number of 
elements not analysed by previous studies.
We also report on the first detection of anomalous shapes of line profiles belonging
to Mn and Hg.
We suggest that the variability of the line profiles of certain elements is caused by a non-uniform surface  distribution of 
these elements similar to the presence of chemical spots detected in HgMn stars.

One important task for future studies is to obtain detailed information on 
the pulsation behaviour of spectral lines in a few PGa stars by monitoring the  
behaviour of line profiles based on 
high quality spectroscopic time series over several hours using short exposures of the order of minutes. 
Current stellar models do not predict nonradial 
pulsations of the order of 5--20\,min in mid B-type stars similar to those detected in roAp stars.
On the other hand, such pulsations were not originally predicted for roAp stars 
either. 
In roAp stars, the pulsation variability is best detected in the line profiles of 
doubly ionised rare earth elements that build element clouds in high atmospheric layers.
In close parallel to roAp stars, a theoretical consideration of B-type stars 
with Hg and Mn overabundances
suggest that these elements can be concentrated in high-altitude clouds (above $\log\tau=-4$;
e.g., \citealt{michaud1974}; \citealt{alecian2011}), with a potential effect of weak magnetic fields on 
their formation (e.g.\ \citealt{alecian2013}).
Clearly, a careful investigation of the variability of mid B-type stars 
is of great interest to studies of stellar structure and evolution.

Our measurements of the magnetic field with the moment technique using 22~\ion{Mn}{ii} lines  
indicate the potential presence of a weak variable longitudinal magnetic field 
of the order of tens  of gauss. 
The question of the presence of weak magnetic fields in stars with Hg and Mn overabundances is still under debate.
\citet{bag2012} used the ESO FORS\,1 pipeline
to reduce the full content of the FORS\,1 archive, among them one polarimetric observation of HD\,19400 at MJD\,=\,52852.371.
While \citet{Hubrig2006b} reported for this epoch a mean longitudinal magnetic field 
$\left<B_{\rm z}\right>_{\rm all}=151\pm46$\,G measured using the whole spectrum and a 
longitudinal magnetic field 
$\left<B_{\rm z}\right>_{\rm hyd}=217\pm65$\,G using only the hydrogen lines, \citet{bag2012}
measured $\left<B_{\rm z}\right>_{\rm all}=124\pm85$\,G, i.e.\ a field at a significance level
of only 1.5$\sigma$.
The authors state that very small instrument flexures, negligible in most
of the instrument applications, may be responsible for some spurious
magnetic field detections. 

Our results using high-resolution spectropolarimetry 
are indicative of the potential presence of a weak magnetic field in HD\,19400. 
Applying the moment technique to Mn lines, we measure  a weak negative longitudinal magnetic field 
$\left<B_z\right>=-70\pm23$\,G at 3$\sigma$ level on the first epoch.
At the same epoch the results from the SVD analysis of the observations
using \ion{Mn}{ii} lines show the longitudinal
magnetic field $\left<B_z\right>=-76\pm25$\,G, and for the SVD analysis 
of \ion{Fe}{ii} lines we obtain 
 $\left<B_z\right>=-91\pm35$\,G. However, the obtained FAP values, 0.008 and 0.003, are above the value 
$10^{-3}$, and thus too high for a marginal detection according to \citet{Donati1992}. 
We note that the presented work is based on spectra obtained only at three different nights. 
To get a better insight into the nature of PGa stars,
it is important to carry out a more complete
study based on spectropolarimetric monitoring over the rotation period.

\section*{Acknowledgments}
We thank the referee Gautier Mathys for his useful comments.
Based on observations made with ESO telescopes at the La Silla Paranal Observatory under programme
IDs 71.D-0308(A) and 091.D-0759(A),
and data obtained from the ESO Science Archive Facility under request number MSCHOELLER51580.
This work has made use of the VALD database, operated at Uppsala University,
the Institute of Astronomy RAS in Moscow, and the University of Vienna.


\appendix

\section{Line lists and abundances}

In Table~\ref{tab:line_abundance} we list all lines used in our abundance analysis.

\begin{table*}
\caption[ ]{
Line by line abundances of HD\,19400 from the ATLAS12 model with parameters
$\teff=13\,500$\,K, $\logg=3.9$. In the second and third columns, we give the oscillator strength with the 
corresponding data base source. The low excitation potential is listed in column 4, followed by the equivalent width and 
the derived abundance. For a number of lines, the abundance was derived from  line profiles. 
}
\label{tab:line_abundance}
\font\grande=cmr7
\grande
\centering
\begin{tabular}{lllrrcl}
\hline\noalign{\smallskip}
\multicolumn{7}{c}{HD\,19400[13500,3.9,ATLAS12]}
\\
\hline\noalign{\smallskip}
\multicolumn{1}{c}{$\lambda$(\AA{})} &
\multicolumn{1}{c}{$\log\,gf$}&
\multicolumn{1}{c}{Ref.$^{a}$}&
\multicolumn{1}{c}{$\chi_{\rm low}$}&
\multicolumn{1}{c}{W(m{\AA{}})}&
\multicolumn{1}{c}{$\log$(N$_{\rm elem}$/N$_{\rm tot}$)}&
\multicolumn{1}{l}{Remarks}
\\
\hline\noalign{\smallskip}
\multicolumn{7}{c}{$\log($N(\ion{He}{i})/N$_{\rm tot}) = -2.17\pm0.08$:}\\
\hline\noalign{\smallskip}
3867.4723 & $-$2.038 & NIST5 & 169086.766 & profile & $-$2.155   & \\
3867.4837 & $-$2.260 & NIST5 & 169086.843 & profile & $-$2.155   & \\
3867.6315 & $-$2.737 & NIST5 & 169087.631 & profile & $-$2.155   & \\
4009.2565 & $-$1.447 & NIST5 & 171134.897 & profile & $-$2.155   & \\   
4026.1844 & $-$2.628 & NIST5 & 169086.766 & profile & $-$2.155:: & no fit \\
4026.1859 & $-$1.453 & NIST5 & 169086.766 & profile & $-$2.155:: & \\
4026.1860 & $-$0.704 & NIST5 & 169086.766 & profile & $-$2.155:: & \\
4026.1968 & $-$1.453 & NIST5 & 169086.843 & profile & $-$2.155:: & \\
4026.1983 & $-$0.976 & NIST5 & 169086.843 & profile & $-$2.155:: & \\
4026.3570 & $-$1.328 & NIST5 & 169087.831 & profile & $-$2.155:: & \\
4120.8108 & $-$1.723 & NIST5 & 169086.766 & profile & $-$2.155   & \\
4120.8237 & $-$1.945 & NIST5 & 169086.843 & profile & $-$2.155   & \\
4120.9916 & $-$2.422 & NIST5 & 169087.831 & profile & $-$2.155   & \\
4143.7590 & $-$1.201 & NIST5 & 171134.897 & profile & $-$2.155   & \\
4387.9291 & $-$0.887 & NIST5 & 171134.897 & profile & $-$2.155:  & no fit in red wing \\
4437.5534 & $-$2.015 & NIST5 & 171134.897 & profile & $-$2.00    & \\
4471.4704 & $-$2.211 & NIST5 & 169086.766 & profile & $-$2.155   & no fit in the core \\
4471.4741 & $-$1.036 & NIST5 & 169086.766 & profile & $-$2.155   & \\
4471.4743 & $-$0.287 & NIST5 & 169086.766 & profile & $-$2.155   & \\
4471.4856 & $-$1.035 & NIST5 & 169086.843 & profile & $-$2.155   & \\
4471.4893 & $-$0.558 & NIST5 & 169086.843 & profile & $-$2.155   & \\
4471.6832 & $-$0.910 & NIST5 & 169087.831 & profile & $-$2.155   & \\
4713.1382 & $-$1.276 & NIST5 & 169086.766 & profile & $-$2.155   & \\
4713.1561 & $-$1.499 & NIST5 & 169086.483 & profile & $-$2.155   & \\
4713.3757 & $-$1.976 & NIST5 & 169087.831 & profile & $-$2.155   & \\
4921.9310 & $-$0.443 & NIST5 & 171134.897 & profile & $-$2.301   & no fit in red wing \\
5015.6780 & $-$0.820 & NIST5 & 166277.440 & profile & $-$2.155   & \\ 
5047.7385 & $-$1.587 & NIST5 & 171134.897 & profile & $-$2.155   & \\ 
5875.5987 & $-$1.516 & NIST5 & 169086.766 & profile & $-$2.301   & \\
5875.6139 & $-$0.339 & NIST5 & 169086.766 & profile & $-$2.301   & \\
5875.6148 & $+$0.409 & NIST5 & 169086.766 & profile & $-$2.301   & \\
5875.6251 & $-$0.339 & NIST5 & 169086.843 & profile & $-$2.301   & \\
5875.6403 & $+$0.138 & NIST5 & 169086.843 & profile & $-$2.301   & \\
5875.9663 & $-$0.214 & NIST5 & 169087.831 & profile & $-$2.301   & \\
6678.1517 & $+$0.329 & NIST5 & 171134.897 & profile & $-$2.301   & \\
\hline\noalign{\smallskip}
\multicolumn{7}{c}{$\log($N(\ion{C}{ii})/N$_{\rm tot}) = -4.12\pm0.02$}\\
\hline\noalign{\smallskip}
3918.968 & $-$0.533 & NIST5 & 131724.37 & profile & $-$4.1  & blend \\
4267.001 & $+$0.563 & NIST5 & 145549.27 & profile & $-$4.1  & blend \\
4267.261 & $+$0.716 & NIST5 & 145550.70 & profile & $-$4.1  & \\
4267.261 & $-$0.584 & NIST5 & 145550.70 & profile & $-$4.1  & \\
6578.052 & $-$0.021 & NIST5 & 116537.65  & profile & $-$4.15 & \\
\hline\noalign{\smallskip}
\multicolumn{7}{c}{$\log($N(\ion{O}{i})/N$_{\rm tot}) = -3.9$}\\
\hline\noalign{\smallskip}
6155.971 & $-$1.011 & NIST5 & 86625.757 & profile & $-$3.9 & \\
6155.989 & $-$1.120 & NIST5 & 86625.757 & profile & $-$3.9 & \\
6156.755 & $-$0.899 & NIST5 & 86627.778 & profile & $-$3.9 & \\
6156.778 & $-$0.694 & NIST5 & 86627.778 & profile & $-$3.9 & \\
\hline
\noalign{\smallskip}
\end{tabular}
\end{table*}

\addtocounter{table}{-1}

\begin{table*}
\caption[ ]{Continued.}
\font\grande=cmr7
\grande
\centering
\begin{tabular}{lllrrcl}
\hline\noalign{\smallskip}
\multicolumn{7}{c}{HD\,19400[13500,3.9,ATLAS12]}
\\
\hline\noalign{\smallskip}
\multicolumn{1}{c}{$\lambda$(\AA{})}&
\multicolumn{1}{c}{$\log\,gf$}&
\multicolumn{1}{c}{Ref.$^{a}$}&
\multicolumn{1}{c}{$\chi_{\rm low}$}&
\multicolumn{1}{c}{W(m\AA{})}&
\multicolumn{1}{c}{$\log$(N$_{\rm elem}$/N$_{\rm tot}$)}&
\multicolumn{1}{l}{Remarks}
\\
\hline\noalign{\smallskip}
\multicolumn{7}{c}{$\log($N(\ion{Ne}{i})/N$_{\rm tot}) = -3.77\pm0.07$}\\
\hline\noalign{\smallskip}
5852.488 & $-$0.455 & NIST5 & 135888.717 & 14.1 & $-$3.79 & \\       
6096.163 & $-$0.297 & NIST5 & 134458.287 & 17.7 & $-$3.85 & \\       
6143.063 & $-$0.098 & NIST5 & 134041.840 & 20.9 & $-$3.86 & \\       
6266.495 & $-$0.357 & NIST5 & 134810.740 & 19.6 & $-$3.68 & \\       
6402.248 & $+$0.345 & NIST5 & 134041.840 & 36.7 & $-$3.72 & \\       
6717.043 & $-$0.356 & NIST5 & 135888.717 & 15.3 & $-$3.70 & \\       
\hline\noalign{\smallskip}
\multicolumn{7}{c}{$\log($N(\ion{Na}{i})/N$_{\rm tot}) = -5.71$}\\
\hline\noalign{\smallskip}
5889.950 & $+$0.108 & NIST5 & 0.00 & profile & $-$5.71 & blend \\
5895.924 & $-$0.194 & NIST5 & 0.00 & profile & $-$5.71 & blend \\
\hline\noalign{\smallskip}
\multicolumn{7}{c}{$\log($N(\ion{Mg}{ii})/N$_{\rm tot}) = -5.06$ }\\
\hline\noalign{\smallskip}
4390.572 & $-$0.523 & NIST5 & 80650.020 & profile & $-$5.06 & blend \\
4390.514 & $-$1.478 & NIST5 & 80650.020 & profile & $-$5.06 & blend \\
4427.994 & $-$1.208 & NIST5 & 80619.500 & profile & $-$5.06 & blend \\
4481.126 & $+$0.749 & NIST5 & 71490.190 & profile & $-$5.06 & blend \\
4481.150 & $-$0.553 & NIST5 & 71490.190 & profile & $-$5.06 & blend \\
4481.325 & $+$0.594 & NIST5 & 71491.063 & profile & $-$5.06 & blend  \\
\hline\noalign{\smallskip}
\multicolumn{7}{c}{$\log($N(\ion{Al}{ii})/N$_{\rm tot}) \le -6.77$}\\
\hline\noalign{\smallskip}
4663.046 & $-$0.290 & NIST5 &  85481.35 & not obs & $\le$$-$6.77 & blend \\
5593.300 & $+$0.410 & NIST5 & 106920.56 & not obs & $\le$$-$6.77 & blend \\
\hline\noalign{\smallskip}
\multicolumn{7}{c}{$\log($N(\ion{Si}{ii})/N$_{\rm tot}) = -4.36\pm0.17$}\\
\hline\noalign{\smallskip}
3856.018 & $-$0.406 & NIST5 &  55325.18 & 123.7 & $-$4.63 & \\
3862.595 & $-$0.757 & NIST5 &  55309.35 & 114.7 & $-$4.42 & \\
4075.452 & $-$1.400 & NIST5 &  79355.02 &  22.6 & $-$4.48 & \\
5688.817 & $+$0.126 & NIST5 & 114414.58 &  12.4 & $-$4.40 & blend telluric \\
5701.37  & $-$0.057 & NIST5 & 114327.15 &   8.7 & $-$4.42 & blend telluric \\
5978.93  & $+$0.084 & NIST5 &  81251.32 &  71.0 & $-$4.52 & \\
6347.109 & $+$0.149 & NIST5 &  65500.47 & 164.7 & $-$4.28 & \\
6371.371 & $-$0.082 & NIST5 &  65500.47 & 131.3 & $-$4.34 & \\
6660.532 & $+$0.162 & NIST5 & 116978.38 &  15.5 & $-$4.06 & \\
6671.840 & $+$0.409 & NIST5 & 117178.06 &  19.7 & $-$4.09 & \\
\hline\noalign{\smallskip}
\multicolumn{7}{c}{$\log($N(\ion{Si}{iii})/N$_{\rm tot}) = -4.37\pm0.02$}\\
\hline\noalign{\smallskip}
4552.622 & $+$0.292 & NIST5 & 153377.050 & 16.7 & $-$4.39 & \\
4567.840 & $+$0.068 & NIST5 & 153377.050 & 13.4 & $-$4.35 & \\
\hline
\noalign{\smallskip}
\end{tabular}
\end{table*}

\addtocounter{table}{-1}
\begin{table*}
\caption[ ]{Continued.}
\font\grande=cmr7
\grande
\centering
\begin{tabular}{lllrrcl}
\hline\noalign{\smallskip}
\multicolumn{7}{c}{HD\,19400[13500,3.9,ATLAS12]}
\\
\hline\noalign{\smallskip}
\multicolumn{1}{c}{$\lambda$(\AA{})}&
\multicolumn{1}{c}{$\log\,gf$}&
\multicolumn{1}{c}{Ref.$^{a}$}&
\multicolumn{1}{c}{$\chi_{\rm low}$}&
\multicolumn{1}{c}{W(m\AA{})}&
\multicolumn{1}{c}{$\log$(N$_{\rm elem}$/N$_{\rm tot}$)}&
\multicolumn{1}{l}{Remarks}
\\
\hline\noalign{\smallskip}
\multicolumn{7}{c}{$\log($N(\ion{P}{ii})/N$_{\rm tot}) = -4.26\pm0.15$}\\
\hline\noalign{\smallskip}
4044.576 & $+$0.669 & K12   & 107360.25 &  68.5 & $-$4.43 & \\
4420.717 & $-$0.330 & NIST5 &  88893.22 &  52.9 & $-$4.40 & \\
4452.472 & $-$0.083 & K12   & 105302.37 &  34.0 & $-$4.30 & \\
4463.027 & $+$0.164 & K12   & 105549.67 &  38.2 & $-$4.46 & \\ 
4466.140 & $-$0.560 & NIST5 & 105549.67 &  21.1 & $-$4.20 & reversal ? \\
4475.270 & $+$0.440 & NIST5 & 105549.67 &  48.0 & $-$4.51 & \\
4499.230 & $+$0.470 & NIST5 & 107922.93 &  58.0 & $-$4.16 & \\
4530.823 & $+$0.074 & K12N  & 105302.37 &  37.0 & $-$4.40 & \\
4554.854 & $-$0.084 & K12   & 106001.25 &  33.9 & $-$4.20 & \\
4565.287 & $-$0.520 & NIST5 & 106001.25 &  20.9 & $-$4.23 & \\
4581.716 & $-$1.121 & K12   & 101635.69 &  15.5 & $-$4.01 & \\
4589.846 & $+$0.400 & NIST5 & 103165.61 &  55.2 & $-$4.30 & \\
4602.069 & $+$0.740 & NIST5 & 103667.86 &  62.7 & $-$4.44 & \\
4626.708 & $-$0.320 & NIST5 & 103339.14 &  29.8 & $-$4.23 & \\
4658.309 & $-$0.320 & NIST5 & 103667.86 &  21.0 & $-$4.49 & asymm ?, uncertain laboratory wavelength? \\
4679.028 & $-$0.319 & K12N  & 106001.25 &  19.3 & $-$4.45 & \\
4927.197 & $-$0.799 & K12N  & 103165.61 &  12.5 & $-$4.33 & \\
4935.631 & $-$0.161 & NIST5 & 111507.66 &  17.6 & $-$4.37 & \\
4943.497 & $+$0.060 & NIST5 & 103667.86 &  39.2 & $-$4.27 & \\
5344.729 & $-$0.280 & K12N  &  86597.55 &  55.4 & $-$4.15 & \\
5409.722 & $-$0.390 & NIST5 &  86743.96 &  60.5 & $-$4.14 & \\
5425.880 & $+$0.288 & K12N  &  87124.60 & 103.7 & $-$4.08 & \\
5499.697 & $-$0.441 & K12N  &  87124.60 &  49.1 & $-$4.30 & \\
5541.139 & $-$0.515 & K12N  & 105302.37 &  16.8 & $-$4.23 & \\
6024.178 & $+$0.198 & K12N  &  86743.96 & 106.5 & $-$3.84 & \\
6034.039 & $-$0.151 & K12N  &  86597.55 &  73.5 & $-$4.00 & \\
6043.084 & $+$0.416 & NIST5 &  87124.60 & 103.4 & $-$4.07 & \\
6055.504 & $+$0.056 & NIST5 & 107922.93 &  23.9 & $-$4.27 & \\
6087.837 & $-$0.346 & NIST5 &  86743.96 &  56.7 & $-$4.13 & \\
6165.600 & $-$0.341 & NIST5 &  87124.60 &  49.8 & $-$4.25 & \\
6232.297 & $-$1.652 & K12N  &  87124.60 &   7.8 & $-$4.35 & \\
6435.282 & $-$1.043 & K12   &  87804.10 &  21.2 & $-$4.24 & \\
6713.283 & $-$1.257 & K12   &  86743.96 &  18.8 & $-$4.21 & \\
\hline\noalign{\smallskip}
\multicolumn{7}{c}{$\log($N(\ion{P}{iii})/N$_{\rm tot}) = -4.44\pm0.09$}\\
\hline\noalign{\smallskip}
4059.312 & $-$0.236 & K13Ph & 116885.87 & 22.5 & $-$4.58 & \\
4080.089 & $-$0.494 & K13Ph & 116874.56 & 21.9 & $-$4.35 & \\
4222.198 & $+$0.218 & K13Ph & 117835.95 & 30.8 & $-$4.44 & \\
4246.720 & $-$0.120 & NIST5 & 117835.95 & 23.3 & $-$4.40 & \\
\hline\noalign{\smallskip}
\multicolumn{7}{c}{$\log($N(\ion{S}{ii})/N$_{\rm tot}) = -5.82$}\\
\hline\noalign{\smallskip}
4162.665 & $+$0.777 & NIST5 & 128559.160 & 6.2 & $-$5.82 & \\
\hline\noalign{\smallskip}
\multicolumn{7}{c}{$\log($N(\ion{Ca}{ii})/N$_{\rm tot}) = -5.50$:}\\
\hline\noalign{\smallskip}
3933.663 & $+$0.135 & NIST5 & 0.000 & profile& $-$5.50: & \\
\hline
\noalign{\smallskip}
\end{tabular}
\end{table*}

\addtocounter{table}{-1}

\begin{table*}
\caption[ ]{Continued.}
\font\grande=cmr7
\grande
\centering
\begin{tabular}{lllrrcl}
\hline\noalign{\smallskip}
\multicolumn{7}{c}{HD\,19400[13500,3.9,ATLAS12]}
\\
\hline\noalign{\smallskip}
\multicolumn{1}{c}{$\lambda$(\AA{})}&
\multicolumn{1}{c}{$\log\,gf$}&
\multicolumn{1}{c}{Ref.$^{a}$}&
\multicolumn{1}{c}{$\chi_{\rm low}$}&
\multicolumn{1}{c}{W(m\AA{})}&
\multicolumn{1}{c}{$\log$(N$_{\rm elem}$/N$_{\rm tot}$)}&
\multicolumn{1}{l}{Remarks}
\\
\hline\noalign{\smallskip}
\multicolumn{7}{c}{$\log($N(\ion{Ti}{ii})/N$_{\rm tot}) = -6.35\pm0.07$}\\
\hline\noalign{\smallskip}
4163.644 & $-$0.130 & NIST5 & 20891.660 & 24.4 & $-$6.28 & \\       
4290.215 & $-$0.850 & NIST5 &  9395.710 & 18.8 & $-$6.36 & \\       
4294.094 & $-$0.930 & NIST5 &  8744.250 & 19.6 & $-$6.29 & \\       
4300.042 & $-$0.442 & NIST5 &  9518.060 & 31.5 & $-$6.40 & \\       
4399.765 & $-$1.190 & NIST5 &  9975.920 & 11.4 & $-$6.27 & \\       
4443.801 & $-$0.717 & NIST5 &  8710.440 & 27.1 & $-$6.28 & \\       
4501.270 & $-$0.767 & NIST5 &  8997.710 & 21.2 & $-$6.38 & \\       
4563.757 & $-$0.690 & NIST5 &  9850.900 & 18.2 & $-$6.51 & \\       
4571.971 & $-$0.317 & NIST5 & 12676.970 & 30.6 & $-$6.37 & \\       
\hline\noalign{\smallskip}
\multicolumn{7}{c}{$\log($N(\ion{Cr}{ii})/N$_{\rm tot}) = -6.24\pm0.09$}\\
\hline\noalign{\smallskip}
4592.049 & $-$1.217 & NIST5 & 32854.950 &  9.4 & $-$6.21 & \\       
4616.629 & $-$1.291 & NIST5 & 32844.760 &  5.9 & $-$6.36 & \\       
4618.803 & $-$0.860 & SL    & 32854.950 & 10.7 & $-$6.25 & \\       
4634.070 & $-$0.990 & SL    & 32844.760 &  8.1 & $-$6.26 & \\       
4824.127 & $-$0.980 & SL    & 31219.350 & 13.6 & $-$6.10 & \\       
\hline\noalign{\smallskip}
\multicolumn{7}{c}{$\log($N(\ion{Mn}{ii})/N$_{\rm tot}) = -4.94\pm0.18$}\\
\hline\noalign{\smallskip}
4206.367 & $-$1.553 & KSG & 43258.640 & 30.4 & $-$4.78 & hfs, weak reversal \\       
4292.237 & $-$1.544 & KSG & 43394.439 & 23.6 & $-$4.99 & hfs, flat core \\       
4363.255 & $-$2.094 & K11 & 44899.820 & 15.8 & $-$4.61 & hfs, bad continuum \\       
4365.217 & $-$1.328 & K11 & 53017.160 & 11.6 & $-$5.12 & hfs, reversal \\       
4478.637 & $-$0.935 & K11 & 53597.130 & 21.9 & $-$5.09 & hfs, reversal \\       
4518.956 & $-$1.322 & K11 & 53597.130 & 13.0 & $-$5.03 & \\       
\hline\noalign{\smallskip}
\multicolumn{7}{c}{$\log($N(\ion{Fe}{ii})/N$_{\rm tot}) = -3.80\pm0.14$}\\
\hline\noalign{\smallskip}
4122.659 & $-$3.300 & FW06  & 20830.553 & 36.4 & $-$3.73 & \\
4258.148 & $-$3.480 & FW06  & 21812.045 & 29.9 & $-$3.68 & \\
4273.320 & $-$3.300 & FW06  & 21812.045 & 29.7 & $-$3.86 & \\
4286.271 & $-$1.578 & K13Fe & 62171.624 & 18.6 & $-$3.76 & \\ 
4296.566 & $-$2.930 & FW06  & 21812.045 & 45.4 & $-$3.78 & \\
4303.170 & $-$2.610 & FW06  & 21812.045 & 60.6 & $-$3.63 & \\  
4369.400 & $-$3.580 & FW06  & 22409.818 & 24.5 & $-$3.70 & \\
4416.819 & $-$2.600 & FW06  & 22409.818 & 51.3 & $-$3.91 & \\
4449.611 & $-$1.678 & K13Fe & 63948.803 & 10.3 & $-$3.99 & \\ 
4489.176 & $-$2.970 & FW06  & 22810.346 & 39.9 & $-$3.85 & \\
4491.397 & $-$2.640 & FW06  & 23031.283 & 47.9 & $-$3.93 & \\
4507.091 & $-$1.909 & K13Fe & 62689.874 & 11.9 & $-$3.65 & \\ 
4508.280 & $-$2.350 & FW06  & 23031.283 & 62.9 & $-$3.76 & \\
4515.333 & $-$2.360 & FW06  & 22939.351 & 65.3 & $-$3.61 & \\
4520.218 & $-$2.620 & FW06  & 22637.195 & 52.9 & $-$3.82 & \\
4522.628 & $-$1.990 & FW06  & 22939.351 & 69.6 & $-$3.92 & \\
4541.516 & $-$2.970 & FW06  & 23031.283 & 43.3 & $-$3.74 & \\  
4555.887 & $-$2.250 & FW06  & 22810.346 & 63.5 & $-$3.84 & \\
4576.333 & $-$2.920 & FW06  & 22939.351 & 40.5 & $-$3.88 & \\
4582.830 & $-$3.062 & FW06  & 22939.351 & 34.5 & $-$3.90 & \\
4583.829 & $-$1.940 & FW06  & 22637.195 & 85.7 & $-$3.55 & \\ 
4620.513 & $-$3.190 & FW06  & 22810.346 & 22.9 & $-$4.11 & \\  
\hline
\noalign{\smallskip}
\end{tabular}
\end{table*}

\addtocounter{table}{-1}

\begin{table*}
\caption[ ]{Continued.}
\font\grande=cmr7
\grande
\centering
\begin{tabular}{lllrrcl}
\hline\noalign{\smallskip}
\multicolumn{7}{c}{HD\,19400[13500,3.9,ATLAS12]}
\\
\hline\noalign{\smallskip}
\multicolumn{1}{c}{$\lambda$(\AA{})}&
\multicolumn{1}{c}{$\log\,gf$}&
\multicolumn{1}{c}{Ref.$^{a}$}&
\multicolumn{1}{c}{$\chi_{\rm low}$}&
\multicolumn{1}{c}{W(m\AA{})}&
\multicolumn{1}{c}{$\log$(N$_{\rm elem}$/N$_{\rm tot}$)}&
\multicolumn{1}{l}{Remarks}
\\
\hline\noalign{\smallskip}
\multicolumn{7}{c}{$\log($N(\ion{Fe}{ii})/N$_{\rm tot}) = -3.80\pm0.14$ (cont.)}\\
\hline\noalign{\smallskip}
4629.331 & $-$2.257 & FW06   & 22637.195 &  64.6 & $-$3.81 & \\ 
4635.316 & $-$1.476 & K13FeN & 48039.090 &  46.1 & $-$3.75 & \\
4913.296 & $+$0.016 & J07    & 82978.717 &  37.4 & $-$3.63 & \\
4923.921 & $-$1.206 & FW06   & 23317.635 &  97.0 & $-$3.97 & \\
4993.350 & $-$3.680 & FW06   & 22637.195 &  20.2 & $-$3.71 & \\
5001.953 & $+$0.933 & J07    & 82853.704 &  70.2 & $-$3.68 & \\
5018.436 & $-$1.350 & FW06   & 23317.635 & 105.2 & $-$3.67 & \\
5030.632 & $+$0.431 & FW06   & 82978.717 &  45.5 & $-$3.81 & \\
5061.710 & $+$0.204 & K13Fe  & 83136.508 &  39.9 & $-$3.72 & \\

5169.028 & $-$0.870 & FW06   & 23317.635 & 104.3 & $-$4.18 & \\
5247.956 & $+$0.550 & FW06   & 84938.265 &  47.8 & $-$3.70 & \\
5493.831 & $+$0.259 & FW06   & 84685.245 &  33.5 & $-$3.83 & \\
5506.199 & $+$0.923 & J07    & 84863.382 &  60.1 & $-$3.72 & \\
\hline\noalign{\smallskip}
\multicolumn{7}{c}{$\log($N(\ion{Fe}{iii})/N$_{\rm tot}) = -3.82\pm0.10$}\\
\hline\noalign{\smallskip}
4022.330 & $-$2.040 & K10 & 93392.300 &  9.3 & $-$3.74 & \\
4371.345 & $-$3.026 & K10 & 66464.800 &  9.5 & $-$3.86 & \\
4382.501 & $-$2.980 & K10 & 66523.020 & 12.4 & $-$3.72 & \\
4419.596 & $-$2.207 & K10 & 89084.790 & 23.3 & $-$3.98 & \\  
\hline\noalign{\smallskip}
\multicolumn{7}{c}{$\log($N(\ion{Ni}{ii})/N$_{\rm tot}) = -5.84$}\\
\hline\noalign{\smallskip}
4067.031 & $-$1.847 & K10Ni & 32499.530 & 20.9 & $-$5.84 & \\
\hline\noalign{\smallskip}
\multicolumn{7}{c}{$\log($N(\ion{Ga}{ii})/N$_{\rm tot}) = -5.19\pm0.17$}\\
\hline\noalign{\smallskip}
4251.155 & $+$0.350 & RS94 & 113815.885 & profile & $-$4.75  & hfs, blend \\
4254.073 & $-$0.230 & RS94 & 113842.301 & profile & $-$5.10  & hfs \\
4255.720 & $+$0.634 & NKW  & 113842.301 & profile & $-$5.25  & hfs, blend \\
4255.936 & $-$0.320 & NKW  & 113842.301 & profile & $-$5.25  & \\
4261.488 & $-$1.100 & GUES & 113883.193 & profile & $-$5.15  & hfs, blend \\
4262.014 & $+$0.980 & RS94 & 113883.193 & profile & $-$5.15  & hfs \\
4263.141 & $-$0.500 & GUES & 113883.193 & profile & $-$5.25  & hfs \\
5219.658 & $+$0.350 & GUES & 120550.431 & profile & $-$5.20  & blend \\
5338.240 & $+$0.430 & RS94 & 118429.967 & profile & $-$5.40: & bad spectrum \\
5360.402 & $+$0.420 & RS94 & 118518.461 & profile & $-$5.10  & hfs, reversal ? \\
5363.585 & $+$0.060 & GUES & 118518.461 & profile & $-$5.30  & hfs, blend \\
5416.318 & $+$0.640 & RS94 & 118727.870 & profile & $-$5.20  & hfs \\
5421.274 & $-$0.050 & NKW  & 118727.870 & profile & $-$5.15  & hfs \\
6334.07  & $+$1.000 & RS94 & 102944.595 & profile & $-$5.40  & hfs \\
6419.24  & $+$0.570 & RS94 & 102944.595 & profile & $-$5.30  & hfs \\
6455.92  & $-$0.080 & RS94 & 102944.595 & profile & $-$5.30  & hfs, blend \\
\hline\noalign{\smallskip}
\multicolumn{7}{c}{$\log($N(\ion{Sr}{ii})/N$_{\rm tot}) = -9.07$}\\
\hline\noalign{\smallskip}
4077.709 & $+$0.148 & NIST5 & 0.00 & profile & $-$9.07 & \\
\hline\noalign{\smallskip}
\multicolumn{7}{c}{$\log($N(\ion{Xe}{ii})/N$_{\rm tot}) = -4.65\pm0.17$}\\
\hline\noalign{\smallskip}
4603.030 & $+$0.018 & NIST5 & 95064.38 & 29.1 & $-$4.89 & \\
4844.330 & $+$0.491 & NIST5 & 93068.44 & 48.4 & $-$4.66 & \\
5419.155 & $+$0.215 & NIST5 & 95064.38 & 37.8 & $-$4.51 & \\
6036.170 & $-$0.609 & NIST5 & 95396.74 & 17.8 & $-$4.55 & \\
6051.120 & $-$0.252 & NIST5 & 95437.67 & 26.0 & $-$4.45 & self-reversal ? \\
6097.570 & $-$0.240 & NIST5 & 95396.74 & 18.1 & $-$4.88 & \\
\hline
\noalign{\smallskip}
\end{tabular}
\end{table*}

\addtocounter{table}{-1}

\begin{table*}
\caption[ ]{Continued.}
\font\grande=cmr7
\grande
\centering
\begin{tabular}{lllrrcl}
\hline\noalign{\smallskip}
\multicolumn{7}{c}{HD\,19400[13500,3.9,ATLAS12]}
\\
\hline\noalign{\smallskip}
\multicolumn{1}{c}{$\lambda$(\AA{})}&
\multicolumn{1}{c}{$\log\,gf$}&
\multicolumn{1}{c}{Ref.$^{a}$}&
\multicolumn{1}{c}{$\chi_{\rm low}$}&
\multicolumn{1}{c}{W(m\AA{})}&
\multicolumn{1}{c}{$\log$(N$_{\rm elem}$/N$_{\rm tot}$)}&
\multicolumn{1}{l}{Remarks}
\\
\hline\noalign{\smallskip}
\multicolumn{7}{c}{$\log($N(\ion{Hg}{ii})/N$_{\rm tot}) = -6.16\pm0.13$}\\
\hline\noalign{\smallskip}
3983.931 & $-$1.510 & NIST5 &  35514.624 & profile & $-$6.35 & \\
5677.102 & $+$0.820 & NIST5 & 105544.042 & 12.6    & $-$6.07 & \\
6149.470 & $+$0.150 & NIST5 &  95714.406 & profile & $-$6.07 & blend \\
\hline
\noalign{\smallskip}
\end{tabular}
\begin{flushleft}
$^{a}$: NIST5: NIST Atomic Spectra Database, version 5 at http://physics.nist.gov/pml/data/asd.cfm;\\
\ion{P}{ii}: K12: http://kurucz.harvard.edu/atoms/1501/gf1501.pos;
K12N: the NIST5 $\log\,gf$ values are replaced by the K12 values;\\
\ion{P}{iii}: K13Ph: http://kurucz.harvard.edu/atoms/1502/gf1502.pos;\\
\ion{Cr}{ii}: SL: \citet{sig1990};\\
\ion{Mn}{ii}: K11: Kurucz, R.~L.\ 2011, private communication;
KSG: \citet{kling2001};\\
\ion{Fe}{ii}: FW06: \citet{FuhrWiese2006}; J07: S.~Johansson (2007), priv.\ comm.; 
K10: http://kurucz.harvard.edu/atoms/2602/gf2602.pos;\\
K13Fe: http://kurucz.harvard.edu/atoms/2601/gf2601.pos;
K13FeN: the NIST5 $\log\,gf$ values are replaced by the K13Fe values; \\
\ion{Ni}{ii}: K10Ni: http://kurucz.harvard.edu/atoms/2801/gf2801.pos;\\
\ion{Ga}{ii}: NKW: \citet{niels2000}; RS94: \citet{ryab1994};
GUES: Guessed values on the basis of the line intensity.
\end{flushleft}
\end{table*}

\label{lastpage}

\end{document}